\documentclass[superscriptaddress,twocolumn,aps,notitlepage,preprintnumbers,nobibnotes,reprint]{revtex4-1}

\usepackage{amsmath}
\usepackage{physics}
\usepackage{amsfonts}
\usepackage{graphicx}
\usepackage[dvipsnames]{xcolor}
\usepackage{qcircuit}
\usepackage{dsfont}
\usepackage{xfrac}
\usepackage{xr}
\usepackage{enumitem}
\usepackage{mathtools}
\usepackage{geometry}

\newcommand{\thetitle}{Programming a quantum computer with quantum instructions}

\newcommand{\beq}{\begin{equation}}
\newcommand{\eeq}{\end{equation}}

\newcommand{\DMEorg}{\ensuremath{\textsf{DME}}}
\newcommand{\DMEsqm}{\ensuremath{\textsf{DME}_2}}
\newcommand{\DMEqme}{\ensuremath{\textsf{DME}_2}}

\newcommand{\SWAP}{\textsf{SWAP}}
\newcommand{\SQM}{\textsf{QME}}
\newcommand{\QME}{\textsf{QME}}
\newcommand{\SQMsubscript}{\nu}

\newcommand{\DME}{\textsf{DME}}
\newcommand{\Id}{\mathds{1}}
\newcommand{\RX}[1]{\textsf{R}_X( #1 )}

\newcommand{\RZ}[1]{\textsf{R}_Z( #1 )}

\newcommand{\targinit}{\sigma_\text{in}}
\newcommand{\ctrlinit}{\rho_\text{in}}

\newcommand{\sigideal}{\sigma_\text{ideal}}
\newcommand{\sigsimideal}{\sigma_{\DMEsqm}}
\newcommand{\id}{\text{ideal}}

\newcommand{\DM}[1]{|#1\rangle\langle #1|}

\newcommand{\Nopt}{N_\text{opt}}

\makeatletter
\newcommand*{\shifttext}[2]{%
  \settowidth{\@tempdima}{#2}%
  \makebox[\@tempdima]{\hspace*{#1}#2}%
}
\makeatother

\newcommand{\RLEaffil}{Research Laboratory of Electronics, Massachusetts Institute of Technology, Cambridge, MA, USA}
\newcommand{\EECSaffil}{Department of Electrical Engineering \& Computer Science, Massachusetts Institute of Technology, Cambridge, MA, USA}
\newcommand{\LLaffil}{MIT Lincoln Laboratory, Lexington, MA, USA}
\newcommand{\CHALMERSaffil}{Microtechnology and Nanoscience, Chalmers University of Technology, G\"oteborg, Sweden}
\newcommand{\MITPHYSaffil}{Department of Physics, Massachusetts Institute of Technology, Cambridge, MA, USA}
\newcommand{\MITMECHaffil}{Department of Mechanical Engineering, Massachusetts Institute of Technology, Cambridge, MA, USA}
\newcommand{\DUKEaffil}{Departments of Physics \& Electrical and Computer Engineering, Duke University, Durham, NC, USA}

\begin{document}
\date{\today}

\title{\thetitle}

% Authors:
\author{M.~Kjaergaard$^\dagger$}\email{mortenk@mit.edu}\affiliation{\RLEaffil}
\author{M.~E.~Schwartz}\thanks{Authors contributed equally to this work}\affiliation{\LLaffil}
\author{A.~Greene}\affiliation{\RLEaffil}
\author{G.~O.~Samach}\affiliation{\RLEaffil}\affiliation{\LLaffil}
\author{A.~Bengtsson}\affiliation{\RLEaffil}\affiliation{\CHALMERSaffil}
\author{M.~O'Keeffe}\affiliation{\LLaffil}
\author{C.~M.~McNally}\affiliation{\RLEaffil}
\author{J.~Braum\"uller}\affiliation{\RLEaffil}
\author{D.~K.~Kim}\affiliation{\LLaffil}
\author{P.~Krantz}\thanks{Current address: Microtechnology and Nanoscience, Chalmers University of Technology}\affiliation{\RLEaffil}
\author{M.~Marvian}\affiliation{\RLEaffil}\affiliation{\MITMECHaffil}
\author{A.~Melville}\affiliation{\LLaffil}
\author{B.~M.~Niedzielski}\affiliation{\LLaffil}
\author{Y.~Sung}\affiliation{\RLEaffil}
\author{R.~Winik}\affiliation{\RLEaffil}
\author{J.~Yoder}\affiliation{\LLaffil}
\author{D.~Rosenberg}\affiliation{\LLaffil}
\author{K.~Obenland}\affiliation{\LLaffil}
\author{S.~Lloyd}\affiliation{\RLEaffil}\affiliation{\MITMECHaffil}
\author{T.~P.~Orlando}\affiliation{\RLEaffil}
\author{I.~Marvian}\affiliation{\DUKEaffil}
\author{S.~Gustavsson}\affiliation{\RLEaffil}
\author{W.~D.~Oliver}\affiliation{\RLEaffil}\affiliation{\LLaffil}\affiliation{\MITPHYSaffil}\affiliation{\EECSaffil}

% Make the title + abstract:
\begin{abstract}
The equivalence between the instructions used to define programs and the input data on which the instructions operate is a basic principle of classical computer architectures and programming~\cite{turing_computable_1937}.
Replacing classical data with quantum states enables fundamentally new computational capabilities with scaling advantages for many applications~\cite{montanaro_quantum_2016,shor_polynomial-time_1999}, and numerous models have been proposed for realizing quantum computation~\cite{divincenzo_quantum_1998,farhi_quantum_2000,raussendorf_one-way_2001}.
However, within each of these models, the quantum data are transformed by a set of gates that are compiled using solely classical information.
Conventional quantum computing models thus break the instruction-data symmetry: classical instructions and quantum data are not directly interchangeable.
In this work, we use a density matrix exponentiation protocol~\cite{lloyd_quantum_2014} to execute quantum instructions on quantum data.
In this approach, a fixed sequence of classically-defined gates performs an operation that uniquely depends on an auxiliary quantum instruction state.
Our demonstration relies on a 99.7\% fidelity controlled-phase gate implemented using two tunable superconducting transmon qubits, which enables an algorithmic fidelity surpassing 90\% at circuit depths exceeding 70.
The utilization of quantum instructions obviates the need for costly tomographic state reconstruction and recompilation, thereby enabling exponential speedup for a broad range of algorithms, including quantum principal component analysis~\cite{lloyd_quantum_2014}, the measurement of entanglement spectra~\cite{pichler_measurement_2016}, and universal quantum emulation~\cite{marvian_universal_2016}.
\end{abstract}

\maketitle

% Article:
% 2.000 - 3.000 word
% Separate unreferenced abstract
% Then introparagraph

% -------------------------------------------------------------
%  INTRODUCTION
% -------------------------------------------------------------

\begin{figure*}[!t]
\center
	\includegraphics{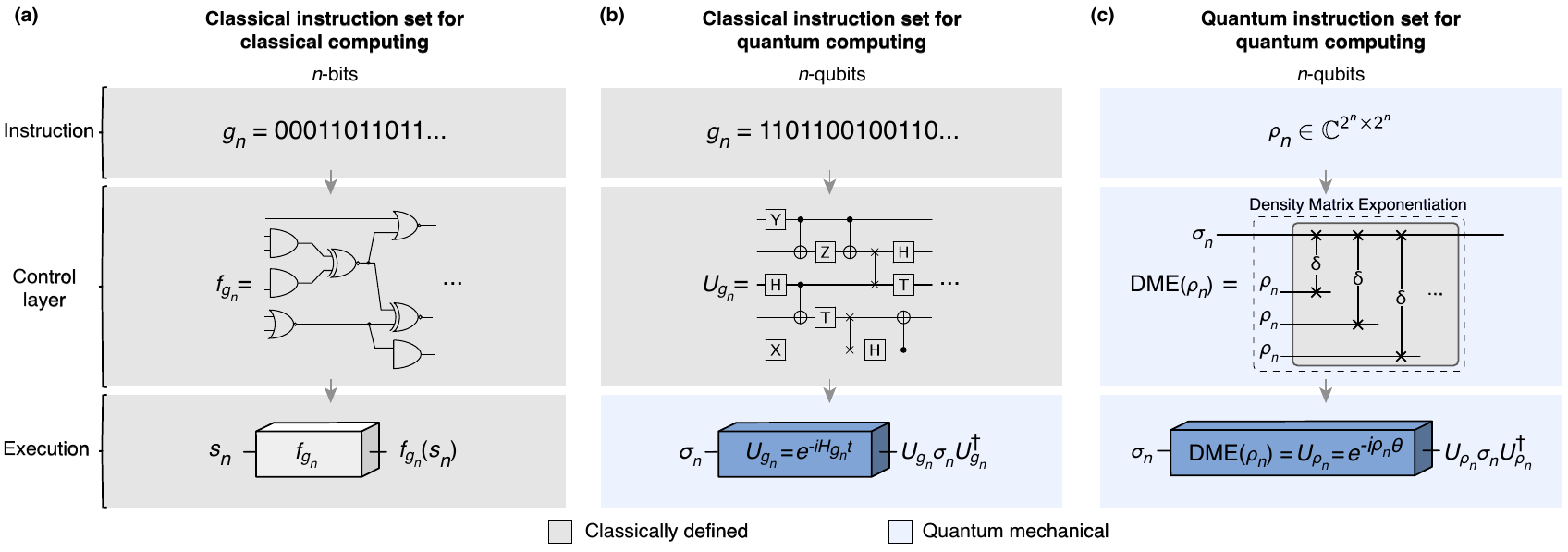}
		\caption{\footnotesize{
		\textbf{Schematic representations of computing architectures.}
		\textbf{a}, Classical instruction set for classical computing.
            Instructions are defined by a classical bit string $g_n$ that uniquely determines a boolean-logic function $f_{g_n}$ comprising single-bit and two-bit gates.
            The control layer executes the resulting circuit on data bits $s_n$ to produce the output $f_{g_n}(s_n)$.
        \textbf{b}, Classical instruction set for conventional quantum computing.
		    The instruction set encoding a quantum circuit is generated using classical resources.
            Instructions are defined by a classical bit string $g_n$ that uniquely determines a unitary operation $U_{g_n}$ comprising single-qubit and two-qubit gates.
            The control layer uses solely classical hardware to generate the gate sequence and applies it to the quantum hardware (data qubits $\sigma_n$) to execute the unitary evolution $U =\exp(-i H_{g_n} t)$, where $H_{g_n}$ is the quantum circuit Hamiltonian, to produce the output $U_{g_n} \sigma_n U^{\dagger}_{g_n}$.
		\textbf{c}, Quantum instruction set for quantum computing using the density matrix exponentiation (\DMEorg) algorithm.
            The instruction set encoding a quantum circuit is stored in the instruction qubits $\rho_n$.
            The control layer uses classical hardware to generate $N$ partial \SWAP{} operations (see text) over a small, classically chosen rotation angle $\delta=\theta / N$, where $\theta$ is an algorithm-dependent angle.
            These classically defined operations (grey region) contain no information about the operation implemented on the data qubits ($\sigma_n$).
            Using a Trotterization approach (see text), the partial \SWAP{} operations are repeatedly applied to the quantum hardware (blue region) -- data qubits $\sigma_n$ and identically prepared copies of the instruction qubits $\rho_n$ -- to execute the unitary operation
            $U =\exp(-i \rho_n \theta) \equiv \exp(-i H_{g_n} t)$, for appropriately chosen $g_n$.
            The output $U_{\rho_n} \sigma_n U_{\rho_n}^{\dagger}$ is equivalent to $U_{g_n} \sigma_n U_{g_n}^{\dagger}$
            to within an error $\mathcal O \left( \theta^2/N \right)$ for appropriately chosen $\rho_n$ (see text).
		}}
	\label{fig:DME_cartoon}
\end{figure*}

In classical programmable computers, instructions are specified in the same medium as the data they process (Figure \ref{fig:DME_cartoon}a), such that programs can can be treated interchangeably with data.
This property, known as homoiconicity~\cite{mooers_programming_1965}, is a hardware-level property of all classical computers based on the von Neumann architecture~\cite{von_1945_first} as well as of higher-level programming languages like Lisp, Julia, and Wolfram.

In all previous experimentally-realized quantum computational systems (e.g., Refs.~\cite{arute_quantum_2019,harris_phase_2018,barz_demonstration_2012}), the relation between instructions and data is fundamentally different.
Quantum programs generally comprise a classically-defined list of gates (the instructions) that are applied to a quantum processor (the data) using intermediary control hardware (Figure~\ref{fig:DME_cartoon}b).
Thus, this programming architecture is non-homoiconic: the instructions are manifestly classical, whereas the data are quantum mechanical.
Furthermore, if an algorithm requires instructions derived from the present quantum state of the processor, that information must first be extracted from the quantum system (incurring exponential overhead~\cite{haah_sample-optimal_2017}), classically processed and compiled, and then appended to the classical instruction list~\cite{vijay2012stabilizing,riste2013deterministic,reed2012realization,hu2019quantum,ofek2016extending,Campagne_2013_persistent,Andersen_2019}.
Such a costly reconstruction process would create significant bottlenecks in quantum algorithms at scale~\cite{Preskill_2018,botea2018complexity}.

Here, we demonstrate the use of quantum instructions to implement a quantum program -- a unitary operation whose parameters are given by the properties of a quantum state~\cite{nielsen_programmable_1997,ying_foundations_2016,yang_optimal_2020} -- on a superconducting quantum processor.
Our approach is based on density matrix exponentiation (\DMEorg), a protocol originally introduced in the context of quantum machine learning~\cite{lloyd_quantum_2014}.
%
% The DME protocol
\DMEorg{} is executed by a series of repeated classical control pulses (Fig.~\ref{fig:DME_cartoon}c), which, in contrast to conventional quantum programs (Fig.~\ref{fig:DME_cartoon}b), carry no information about the instruction set.
Rather, the instructions are encoded in the ``instruction qubits'' (density matrix $\rho$), which determine the unitary operations performed on the ``data qubits'' (density matrix $\sigma$).
Thus, both instructions and data are stored in quantum states~\cite{nielsen_programmable_1997}, and together they constitute a quantum computing analogue of homoiconicity.
We apply \DMEorg{} to a system comprising two superconducting qubits: a data qubit prepared in state $\sigma$, and an instruction qubit prepared in state $\rho$. In this case, \DMEorg{} implements a unitary rotation on the data qubit about an axis parallel to instruction state.
At this scale, the quantum instruction state and its resulting unitary operation are easily predicted and straightforward to reconstruct.
However, extending the number of instruction qubits to the supremacy regime~\cite{arute_quantum_2019}, in which a classically-specified sequence of gates can produce a quantum state that is too complex to predict and too large to tomographically reconstruct, leads to a remarkable programming and operational framework: while it would be completely impractical to ascertain the quantum instructions stored in an unknown state $\rho$, one can nonetheless use these instructions to execute a quantum program (as defined above).
Programs based on quantum instructions implemented with \DMEorg{} enable
a class of efficient algorithms addressing both quantum computation (using $\rho$ to manipulate $\sigma$) and quantum metrology (using $\sigma$ to study $\rho$).
The afforded quantum advantage generally stems from the fact that \DMEorg{} directly implements unitary operations $e^{-i\rho\theta}$ at the quantum hardware layer, obviating the need for tomographic reconstruction of the instruction state $\rho$ on which these applications are based~\cite{lloyd_quantum_2014,kimmel_hamiltonian_2017,haah_sample-optimal_2017}.
One example is private quantum software execution, whereby the action of an unknown (private) unitary $U$ on an arbitrary quantum state may be efficiently emulated using a relatively small set of known input-output relations $\{\rho_\text{in}\} \stackrel{U}{\mapsto} \{\rho_\text{out}\}$~\cite{marvian_universal_2016}, far fewer than would be required to compromise the security of $U$ via its tomographic reconstruction.
Quantum instructions also enable quantum advantage for quantum semi-definite programming~\cite{brandao_quantum_2019} and sample-optimal Hamiltonian simulation~\cite{kimmel_hamiltonian_2017}.
In addition, quantum phase estimation executed using \DMEorg{} can extract with error $\epsilon$ the dominant eigenvalues and eigenvectors of $\rho$ -- principal component analysis -- using only $\mathcal O(\theta^2/\epsilon)$ copies of $\rho$~\cite{lloyd_quantum_2014,kimmel_hamiltonian_2017}.
Even when $\rho$ is a large entangled state, \DMEorg{} can efficiently reveal its entanglement spectrum, a form of reduced-complexity benchmarking~\cite{pichler_measurement_2016}.
%
% -------------------------------------------------------------
%  ALGORITHM
% -------------------------------------------------------------
\begin{figure*}[!ht]
\centering
	\includegraphics[width=6.5in]{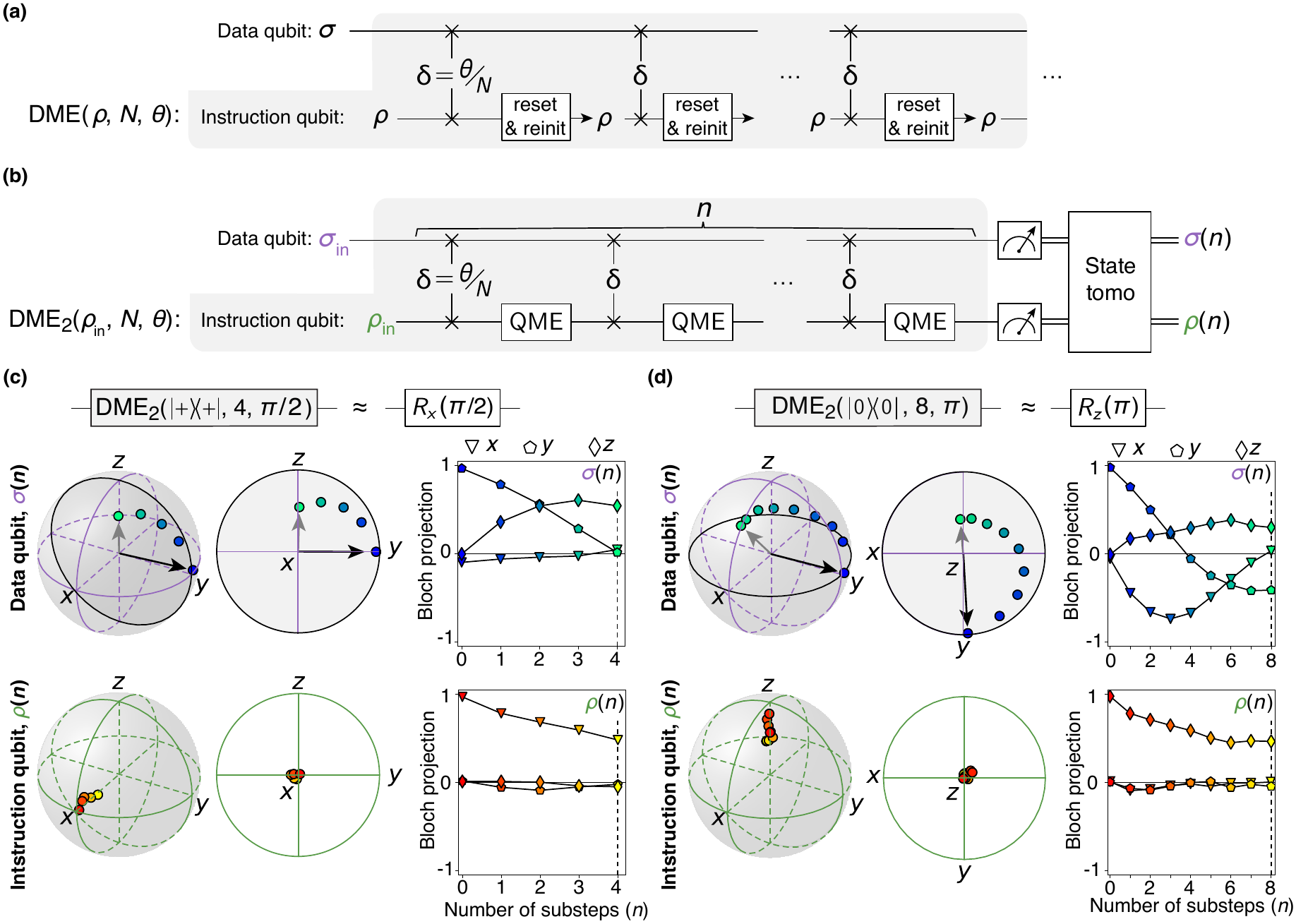}
		\caption{
		\textbf{Demonstration of quantum instructions}.
        \textbf {a} Density Matrix Exponentiation (\DMEorg) algorithm using active reset and reinitialization to reprepare the instruction state $\rho$ after each $\delta\SWAP$ operation.
		\textbf {b} Implementation of \DMEsqm{}, in which quantum measurement emulation (\SQM{}) is used to approximately reinitialize the instruction qubit to $\ctrlinit$ without active reset and repreparation (see text for details). The substep parameter $n$  is stepped from 0 to $N$. In the experiment, we perform $n$ rounds of $\delta\SWAP+\SQM$, measure the two-qubit density matrix, and trace over each subsystem to extract the individual data and instruction qubit density matrices ($\sigma(n)$ and $\rho(n)$ respectively).
		\textbf {c} Substeps of $\DMEsqm(\ketbra{+}, 4, \pi/2)$, corresponding to $\RX{\pi/2}$ on the target qubit at the final step $(n=N)$. Black lines are guides to the eye.
		\textbf {d} Substeps of $\DMEsqm(\ketbra{0}, 8, \pi)$, corresponding to $\RZ{\pi}$ on the target qubit at $n=N$.}
	\label{fig:DME_bloch}
\end{figure*}
\\
\\
\noindent
\textbf{The \DMEorg{} protocol}
\\
Conceptually, \DMEorg{} implements the unitary operation $U=e^{-i\rho\theta}$ on the data qubit(s) according to the instruction state $\rho$ and an algorithm-dependent angle $\theta$.
If the data and instruction states are single-qubit pure states $\sigma$ and $\rho$, \DMEorg{} rotates $\sigma$ by an angle $\theta$ about an axis defined by the Bloch vector of $\rho$.
More generally, $\sigma$ and $\rho$ are multi-qubit states, and they need not be pure states.
The protocol that implements \DMEorg{} partitions $U$ into a sequence of $N$ steps (Fig.~\ref{fig:DME_bloch}a), each comprising a ``partial $\SWAP{}$'' operation $\delta\SWAP~\equiv~e^{-i\SWAP\delta}$~\cite{salathe_digital_2015} that is applied to $\sigma$ and $\rho$~\cite{lloyd_quantum_2014}.
The protocol relies on the relation:
\begin{eqnarray}
	\text{Tr}_\rho\left[e^{-i\SWAP\delta}\sigma \otimes \rho e^{i\SWAP\delta}\right]&=& \sigma - i\delta[\rho,\sigma] + \mathcal O (\delta ^ 2) \nonumber \\
	&=& e^{-i\rho \delta} \sigma e^{i\rho \delta} + \mathcal O (\delta ^ 2).
\end{eqnarray}
That is, $\sigma$ undergoes unitary evolution of the form $e^{-i\rho\delta}$ (to first order in $\delta$, see Supplementary Methods), rotating by a small angle $\delta = \theta / N$.
By the reciprocity of $\SWAP$ operations, $\rho$ undergoes a complementary unitary evolution about $\sigma$, leaving it in a state that differs from the original quantum instruction.
As a result, the instruction qubits must be refreshed at each step to provide a new, identical copy of the instruction state $\rho$.
Repeating these steps $N$ times (Fig.~\ref{fig:DME_bloch}a) approximately yields the desired operator,
\begin{equation}
	\DMEorg(\rho, N, \theta) \rightarrow e^{-i\rho \theta} + \mathcal O \left(\tfrac{\theta^2}{N}\right),
	\label{eq:dme_lmr}
\end{equation}
a result that is closely related to the Trotterization of non-commuting Hamiltonians to perform quantum simulations~\cite{lloyd_universal_1996}.
Similar to dividing a quantum simulation into smaller steps to reduce errors stemming from the Trotter approximation, partitioning $\DMEorg$ into more steps (increasing $N$), with a smaller partial \SWAP{} angle $\delta$ per step, reduces the \DMEorg{} discretization error.
The trade-off for increased precision is a need for more copies of the quantum instructions.
There are three general approaches to supplying the $N$ copies of the instruction state $\rho$ needed to execute $\DMEorg$:
\begin{enumerate}
	\item Teleport copies of the quantum instructions from a third party to the qubits comprising $\rho$.
	\item Identically prepare the instructions on $N$ copies of $\rho$ (hardware parallelization,
            Figure~\ref{fig:DME_cartoon}c);
    \item Identically prepare the same set of qubits comprising $\rho$ after each $\delta$\SWAP{} (sequential preparation in time, Figure~\ref{fig:DME_bloch}a).
\end{enumerate}

In this work, we choose option 3: we refresh the same instruction qubit to avoid the need for teleportation or large numbers of instruction qubits, and to allow us to easily vary $N$.
\\
\\
\textbf{A new approach to generating $N$ copies of $\rho$}
\\
The most obvious approach for using one qubit to generate $N$ copies of $\rho$ (Figure~\ref{fig:DME_bloch}a) is to use measurement-conditioned active feedback to reset the instruction qubit to its ground state and then re-prepare the instruction state $\rho$~\cite{magnard_fast_2018}.
However, due to measurement infidelity and the decoherence that occurs during the relatively long duration of the requisite measurement, feedback, and preparation steps, the conventional active-reset approach would introduce an unacceptable level of errors on current quantum processors.
Here, to minimize such errors and achieve the largest possible circuit depths with our qubits, we instead introduce an alternative approach (Figure~\ref{fig:DME_bloch}b) called quantum measurement emulation (\QME), that approximately reinitializes the instruction qubit in the time required for a single qubit gate.

% -------------------------------------------------------------
%  IMPLEMENTATION DETAILS
% -------------------------------------------------------------
\QME{} is a probabilistic operation that mimics an ensemble averaged qubit measurement.
For intuition, note that for a sufficiently small angle $\delta$, the states of the two qubits are only slightly altered after a $\delta\SWAP$ operation.
In this case, a projective measurement of the instruction qubit in the eigenbasis of $\rho$ would reset the instruction qubit to its original state with high probability.
Similarly, an unconditioned ensemble averaged measurement of many such identically prepared states (i.e. a measure-and-forget approach) would reproduce the original $\rho$ with only a slight depolarization.
\QME{} mimics this well-known result without actually performing a measurement by imposing a dephasing channel corresponding to the axis of $\rho$.
%
%\wdo{Although one would not know the axis of $\rho$ in the general case, leaving no alternative but to use active feedback (Figure~\ref{fig:DME_bloch}a), here we do know the axis of $\rho$ -- by construction -- and can implement \QME{} as a tool without loss of generality of our purpose: a demonstration of quantum instructions via \DME.}

The \QME{} operation (Figure~\ref{fig:DME_bloch}b) randomly applies either an identity gate ($\Id$) or a $\pi$-rotation in the instruction qubit eigenbasis, according to a Bernoulli process with probability $p=0.5$:
\begin{equation}
\Qcircuit @C=0.5em @R=1em{
	&\gate{\QME_{\SQMsubscript}} & \qw\\
}
=
\begin{cases}
\Id & \text{with } p = 0.5\\
\textsf{R}_{{\SQMsubscript}}(\pi) & \text{with } 1-p = 0.5
\end{cases}
\label{eq:sqm_def}
\end{equation}
where ${\SQMsubscript}$ is a normalized vector parallel to the original instruction state.
For instruction states aligned with the axes of the Bloch sphere, \QME{} represents a probabilistic application of a Pauli gate.
We incorporate \QME{} into our circuit by interleaving $\delta\SWAP$ and \QME{} operations (Figure~\ref{fig:DME_bloch}b). 
The \QME{} operations are randomized within each instantiation of the circuit, and in the same spirit as randomized compiling~\cite{wallman_noise_2016}, the outcomes of multiple such randomized instantiations are averaged to mimic a single circuit with an active reset of $\rho$.

The \QME-enabled reset of $\rho$ is approximate due to depolarization, which introduces an additional error term to the $\DMEorg$ protocol.
For our two-qubit demonstration with $\QME$, denoted $\DMEqme$, the unitary operation is (see Methods): 
\begin{equation}\begin{array}{cll}
\DMEqme(\rho, N, \theta)
    & \rightarrow & e^{-i \rho \theta} +\underbrace{\mathcal O(\tfrac{\theta^2}{N})}_{\text{discretization}} +\underbrace{\mathcal O (\tfrac{\theta^2}{N})}_{\QME}.
\end{array}
\label{eq:dme_sqm_circuit}
\end{equation}
Despite the added error term, \QME{} effectively supplies the requisite copies of $\rho$ with less error than would be incurred with an active feedback approach in our system.
Furthermore, its use here is not fundamental to our demonstration.
An implementation of arbitrary, unknown quantum instructions could be performed using states from large quantum processors or by state-teleportation, and the underlying physics would be the same as those in this proof-of-principle experiment.
\\
\\
\\
\textbf{Implementing the \DMEsqm{} algorithm}

We implement \DMEsqm{} using two frequency-tunable superconducting `asymmetric' transmon qubits~\cite{koch_charge-insensitive_2007,hutchings_tunable_2017} in an `xmon' layout~\cite{barends_coherent_2013}, operating with single-qubit gate fidelities exceeding 99.9\% and a two-qubit controlled phase gate~\cite{strauch_quantum_2003,dicarlo_2009} with 99.7\% fidelity  (see Methods).
In Figures~\ref{fig:DME_bloch}c and~\ref{fig:DME_bloch}d, we interrupt the algorithm after $n\leq N$ steps and perform state tomography (see Methods) to visualize the evolution of the data-qubit and instruction-qubit.
We use an initial state $\targinit = \ketbra{+i}$ for the data qubit, and we introduce the notation $\DMEsqm(\ctrlinit, N, \theta)$ to indicate the initial instruction state $\ctrlinit$, total number of steps $N$, and the phase rotation $\theta$.
Figure~\ref{fig:DME_bloch}c shows an implementation of \DMEsqm$(\DM{+}, 4, \pi/2)$. 
Since $\rho$ is $x$-polarized, this instruction encodes the operation $\RX{\pi/2}$, a $\pi/2$ rotation about the $x$ axis. 
Figure~\ref{fig:DME_bloch}d shows an implementation of \DMEsqm$(\DM{0},8,\pi)$, encoding the instruction $\RZ{\pi}$, a $\pi$ rotation about the $z$ axis. 
In both cases, $\sigma$ undergoes a rotation about an axis defined by $\ctrlinit$, which is visible in the step-by-step tomographic reconstruction of the data qubit state $\sigma(n)$.
$\QME{}$ maintains the polarization direction of the instruction qubit state $\rho(n)$, albeit with gradual depolarization consistent with the effects of the simulated measurement.
The classically-defined $\delta\SWAP$ operations are identical in these two cases; it's the change in the instruction state $\ctrlinit$ that causes a different operation on the data qubit.
Thus, the implemented quantum program (here, a single-qubit operation) is uniquely determined by the state of another quantum system, a demonstration of quantum instructions.

\begin{figure}[!t]
\center
	\includegraphics[width=3.2in]{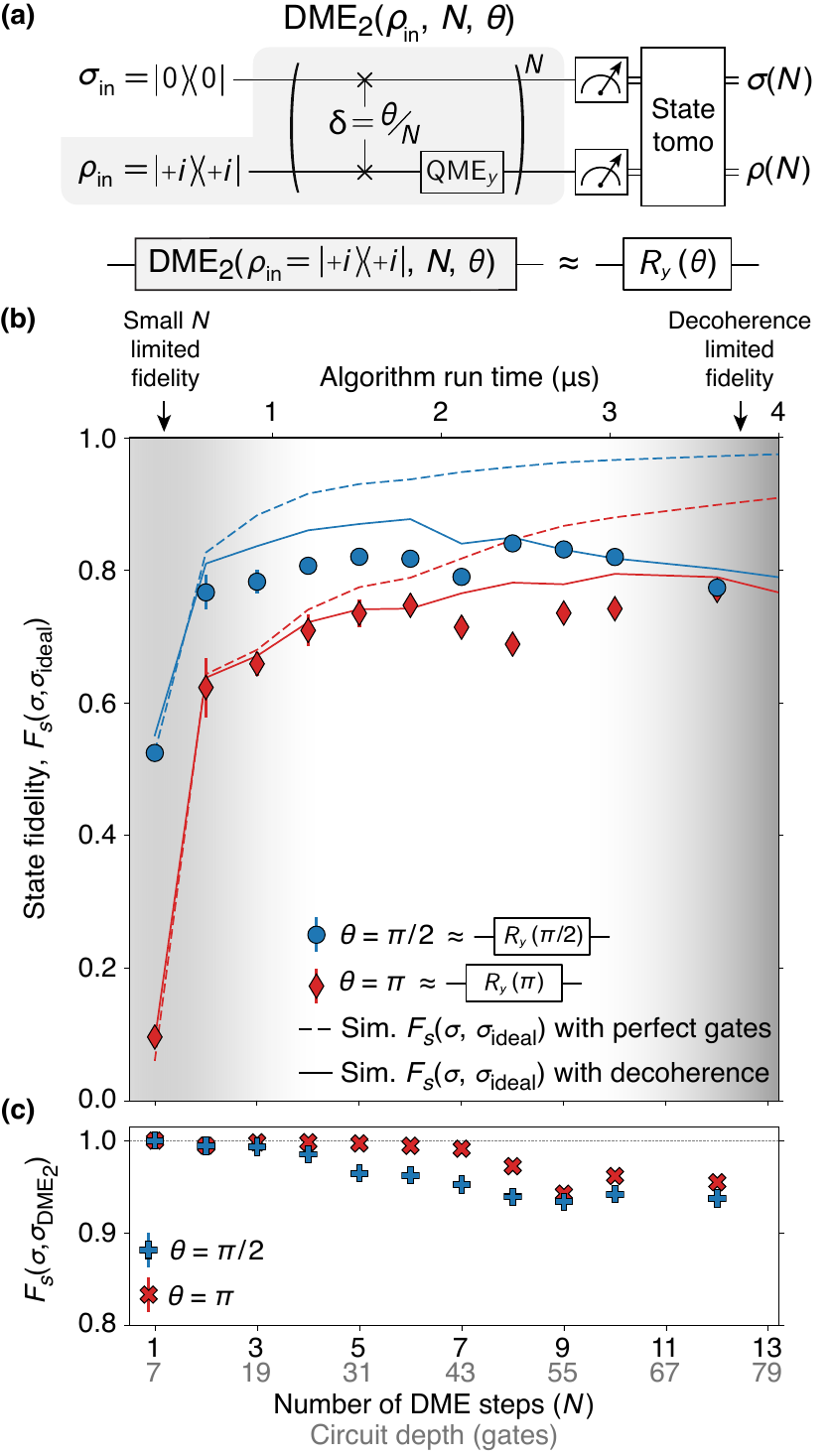}
		\caption{\footnotesize{
		\textbf{Algorithm performance as a function of $N$}.
		\textbf a Circuit schematic for \DMEsqm$(\ketbra{+i}, N, \theta)$. Data qubit is initialized in $\targinit = \ketbra{0}$.
		\textbf b State fidelity ($F_s$) of the data qubit state $\sigma$ to the ideal state $\sigma_\text{ideal}~=~e^{-i\ctrlinit\theta}\targinit e^{i\ctrlinit\theta}$ as a function of total \DME{} steps $(N)$. 
        The instruction qubit is initialized to the $\ketbra{+i}$--state, resulting in an ideal operation $e^{-i\ctrlinit\theta} = R_y(\theta)$.
		The $x$-axis shows the number of $\delta\SWAP + \QME$ steps $N$ (bottom, black), circuit depth (bottom, gray), and active circuit clock time (top).
		Data for $\theta = \pi$ $(\pi/2)$ are shown with red/$\diamond$ (blue/$\circ$) markers. 
        Dashed lines is the state fidelity between $\sigma_\text{ideal}$ and a simulated output of the \DMEsqm$(\ketbra{+i}, N, \theta)$ circuit, assuming perfect gates.
        The increasing fidelity with increasing $N$ is a reflection of a reduction of the `discretization error' scaling as $\mathcal O (\theta^2/N)$.
        Solid lines are the same simulation as shown in dashed lines, but with amplitude damping and depolarizing channels included in the circuit to model the effect of decoherence.
  %       indicate the algorithmic error assuming perfect gates;
		% solid lines represent the simulated \DMEsqm{} performance in the presence of amplitude damping and depolarizing channels.
		\textbf c State fidelity of $\sigma$ to a simulated output of the \DMEsqm$(\ketbra{+i}, N, \theta)$ circuit with perfect gates (denoted $\sigma_{\DMEsqm}(N)$).
		Error bars are determined from bootstrap analysis (see Methods).
		}}
	\label{fig:DME_fidelity}
\end{figure}

% ------------------------------------------
%  ALGORITHM CHARACTERIZATION
% -------------------------------------------------------------
We next assess \DMEsqm{} in the context of an imperfect quantum processor with noise-induced errors in addition to discretization (finite $N$) errors (Figure~\ref{fig:DME_fidelity}).
Here, we fix target state $\targinit = \ketbra{0}$ and instruction state $\ctrlinit = \ketbra{+i}$, and vary total steps $N$.
This allows us to probe the interplay between discretization error (which decreases with $N$) and noise-induced errors (which increases with $N$).
We use two angles, $\theta = \pi$ and  $\theta = \pi/2$, to elucidate the effects of changing the overall angle.
For each experiment, we perform the full algorithm $\DMEsqm (\ctrlinit, N, \theta)$ with many \SQM{} randomizations, tomographically reconstruct the final density matrix $\sigma(N)$, and calculate its fidelity to an ideal rotation with no discretization or processor error, as given by $\sigideal~=~e^{-i\ctrlinit\theta}\targinit e^{i\ctrlinit\theta}$. 
$\sigideal$ can equivalently be thought of as a perfect unitary rotation, or as \DMEorg{} when implemented on an error-free processor with an infinite number of copies of the quantum instructions.

There are two sources of error we must consider in understanding the output of the \DMEsqm{} protocol: the approximate nature of the algorithm and imperfections in the quantum processor.
To understand the discretization error, we calculate $\sigsimideal$, the outcome of a simulation of the \DMEsqm{} circuit (including discretization error) with perfect gates.
We sample every possible combination of \SQM{} gates for a \DMEsqm{} circuit of length $N$ and simulate the application of each circuit to the experimentally-measured $\ctrlinit \otimes \targinit$ (thus accounting for state-preparation errors).
%We average all simulated outcomes and trace over the instruction qubit to generate $\sigsimideal$.
The fidelity $F_s(\sigsimideal, \sigideal)$ quantifies the error due solely to the approximate nature of \DMEsqm{} (Figure~\ref{fig:DME_fidelity}b, dashed lines).
Figure \ref{fig:DME_fidelity}c shows the fidelity of the measured state $\sigma(N)$ to the ideal algorithm performance, $F_s(\sigma, \sigsimideal)$.
To circuit depth 73, this fidelity exceeds 0.90.

We next account for the effects of imperfections in the physical processor by building a model of \DMEsqm{} performance in the presence of processor noise.
To the \DMEsqm{} circuit with perfect gates we add amplitude-damping and dephasing channels with coherence parameters consistent with independent measurements (see Methods).
The fidelity between the model including decoherence effects and $\sigma_\text{ideal}$ is plotted in Figure~\ref{fig:DME_fidelity}b (solid lines), and shows good agreement with experimental data, indicating we are mostly limited by decoherence effects, and not coherent errors in the gates.

Both the simulated and experimental curves reveal an interplay between finite $N$ error and processor error.
At small $N$, the error is dominated by the approximate nature of \DMEsqm{} as given in Eq.~\eqref{eq:dme_sqm_circuit}.
The error is greater for larger $\theta$, consistent with error scaling as $\mathcal O ( \theta^2/N )$.
For large $N$, the discretization error improves and the processor's performance is instead limited by finite gate fidelity; here, the curves for $\theta=\pi$ and $\theta=\pi/2$ begin to converge.
The algorithm is at its most accurate for intermediate $N$, where discretization error is relatively low and the circuit is sufficiently free of compounding physical errors.
This tradeoff (improved performance with increasing circuit depth, until gate fidelities become limiting) is a generic property of Trotterized quantum algorithms on noisy processors in the absence of error-correction protocols~\cite{rines_empirical_2019}.

\begin{figure*}[t]
\center
	\includegraphics[width=6.7in]{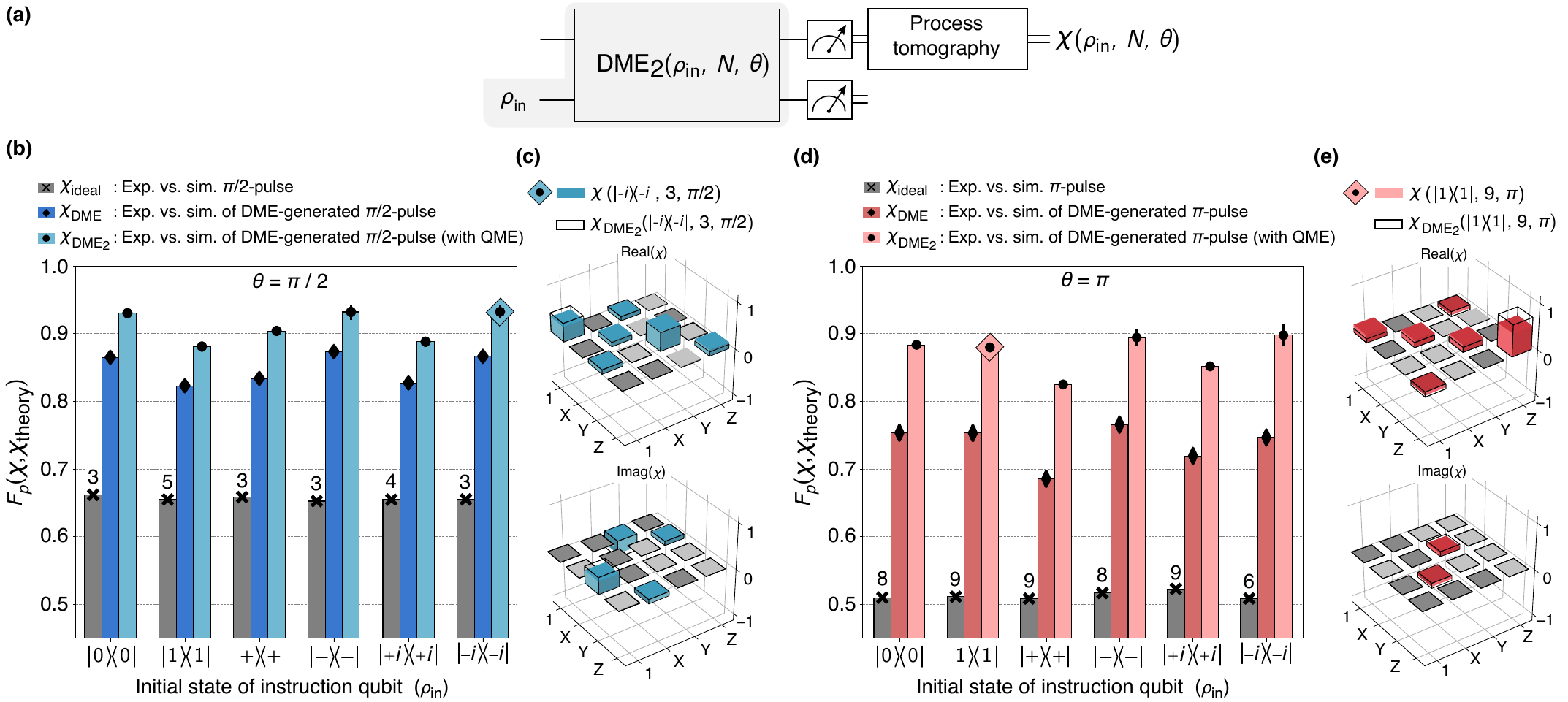}
		\caption{
		\textbf{Benchmarking algorithmic fidelity of \DMEsqm{}}.
		\textbf{a} Circuit schematic. Single-qubit process tomography is performed for a set of six instruction states $\ctrlinit$ representing cardinal points of the Bloch sphere.
		\textbf{b (d)} %
        Process fidelities between measured process maps and simulated processses, for varying instruction state and overall angle in $\DMEsqm$.
		Grey ($\times$ marker) denotes the fidelity $F_p(\chi, \chi_\text{ideal})$ between the measured process map $\chi$ and the ideal process $\chi_\text{ideal}$, e.g a rotation of angle $\theta$ around the axis given by the Bloch vector of $\ctrlinit$.
		The data are presented at $N=N_\text{opt}$, determined as the step-number at which the fidelity to $\chi_\text{ideal}$ is maximized; $N_\text{opt}$ is indicated by the number above each bar.
		Dark blue/red ($\diamond$ marker) indicates the fidelity $F_p(\chi, \chi_{\DMEorg{}})$ between the measured process map and a simulated implementation of the \DMEorg{} circuit assuming active reset and reinitialization of the instruction qubit (evaluated at $N=N_\text{opt}$).
		Light blue/red ($\circ$ marker) shows the fidelity $F_p(\chi, \chi_{\DMEsqm{}})$ between the measured process map and a simulation of $\DMEsqm$ with perfect gates and no decoherence using \SQM{} to approximately reinitialize the instruction qubit at each step.
        %
        % $F_p(\chi, \chi_{\DMEsqm{}})$ represents the fidelity between our experimental $\DMEsqm$ implementation and a simulation with perfect gates and no decoherence of the $\DMEsqm$ circuit.
        %
		Error bars are calculated using bootstrap analysis (see Methods).
		The process map for the point enclosed by a blue/red diamond is shown in \textbf{c (e)}.
        \textbf{c (e)}
        Representative process matrices for $\chi$ shown in blue (red) for $\theta = \pi/2$ ($\theta = \pi$), evaluated at $N_\text{opt}$.
		Colored process matrix elements indicate points with magnitude $\chi_{ij} > 0.02$; other elements are grey for clarity of scale.
		Black wire frames denote a process matrix from a simulated implementaiton of $\chi_{\DMEsqm}$ assuming perfect gate operation and no decoherence.
		}
	\label{fig:DME_processfidelity}
\end{figure*}

To assess the error-budget of the quantum instruction execution independent of its operation on the target quantum state, we perform quantum process tomography.
We experimentally reconstruct the process map of the channel implemented by \DMEsqm{} (denoted $\chi(\ctrlinit, N, \theta)$) for a set of instruction states given by the cardinal points on the Bloch sphere.
For each $\ctrlinit$, we sweep $N$ to find the optimal point $\Nopt$, defined as that which has the highest process fidelity (see Methods) to the pure rotation $U_\id = e^{-i\ctrlinit\theta}$.
The mean $\Nopt$ for $\theta=\pi/2$ is 4, at circuit depth 25; for $\theta=\pi$ this increases to 8, at circuit depth 49.

In Figure \ref{fig:DME_processfidelity} we plot the process fidelity at $\Nopt$ to several theoretical processes to elucidate the error budget in \DMEsqm.
The fidelity to $\chi_\id$, corresponding to the perfect rotation $U_\id$, is plotted in grey.
$F_p(\chi, \chi_\id)$ is greater for $\theta = \pi/2$ than for $\theta = \pi$, as expected from the $\mathcal O (\theta^2/N)$ scaling of the discretization error, and is consistent for all cardinal settings of the instruction state.
This process fidelity reflects the combination of errors arising from the Trotterized nature of density matrix exponentiation and the errors from imperfect gates and the approximate nature of \SQM{}.

We next compare our \SQM-enabled algorithm to the original $\DMEorg$ proposal requiring $N$ copies of $\ctrlinit$.
We label this theoretical process $\chi_{\DMEorg}$ and calculate the fidelity $F_p(\chi, \chi_{\DMEorg})$, shown in dark blue/red.
This fidelity combines the physical errors arising from our imperfect gates and the error from using \SQM{} to emulate the re-preparation of $\ctrlinit$.
The difference between $F_p(\chi, \chi_{\DMEorg})$ and $F_p(\chi, \chi_\text{ideal})$ is a reflection of finite $N$ error in the underlying \DMEorg{} algorithm.

Finally, we plot the fidelity between our measured process and $\chi_{\DMEsqm}$, a simulated version of the \DMEsqm{} algorithm, shown in light blue/red.
This fidelity compares the experimental implementation of $\DMEsqm$ to a classical simulation of \DMEsqm{} using perfect operations, and is therefore the most direct metric for the performance of our processor.
The theoretical $\chi_{\DMEsqm}$ is calculated by sampling all \SQM{} randomizations and averaging their effect.
The average process fidelity $F_p(\chi, \chi_{\DMEsqm})$ over all instruction settings is 0.91 for $\theta = \pi/2$ and 0.87 for $\theta = \pi$; this algorithmic fidelity is overall reduced for $\theta = \pi$ because $N_\text{opt}$ occurs at deeper circuit depth.
% -------------------------------------------------------------
%  CONCLUSION
% -------------------------------------------------------------
\\
\\
\noindent
\textbf{Outlook}
\\
By executing a quantum program whose instructions are stored in a quantum state, we have established the first experimental approach to instilling homoiconicity in quantum computing.
The \DMEsqm{} algorithm used here to generate quantum instructions takes advantage of a 99.7\% fidelity controlled-phase gate combined with a novel simulated quantum measurement technique.
We achieve output state fidelities exceeding 0.9 even at circuit depth of 73 sequential gates and process fidelities close to 0.9, independent of the instruction setting.

Our realization of quantum instructions represents a new approach to quantum computer programming.
It uses a fixed set of classical pulses that serve as a program “scaffolding,” and an auxiliary quantum state to encode the program instructions.
Intuitively, since the classical scaffolding is always the same, it must be the quantum states that uniquely determine the quantum instructions.
While we used pure states to form the quantum instructions in \DMEsqm{}, the general \DMEorg{} algorithm generalizes to mixed states and efficiently extends to multi-qubit systems, requiring only the ability to perform controlled versions of the $\SWAP$ operation between pairs of target and instruction qubits~\cite{gao_entanglement_2019,pichler_measurement_2016,marvian_universal_2016,kimmel_hamiltonian_2017}.
Because it bypasses costly and unscalable tomography and recompilation requirements, \DMEorg{} and related algorithms enable exponential speedup for a range of applications spanning quantum state metrology~\cite{lloyd_quantum_2014}, guaranteed private quantum software~\cite{marvian_universal_2016}, efficient measurement of large-scale entangled quantum systems~\cite{pichler_measurement_2016} and quantum machine learning~\cite{Biamonte2017}.
\\
% By executing a quantum program whose instructions are stored in a quantum state, we have demonstrated an efficent experimental approach to restoring homoiconicity to quantum computing.
% Our implementation relies on a 99.7\% fidelity controlled-phase gate -- a state of the art two-qubit gate -- combined with a novel simulated quantum measurement technique.
% We achieve output state fidelities exceeding $0.9$ even at circuit depth of 73 sequential gates -- to our knowledge one of the deepest circuits executed with high fidelity to date -- and process fidelities close to $0.9$ independent of the instruction setting.
% While we use pure states to form the quantum instruction set, the DME algorithm generalizes to mixed states and efficiently extends to multi-qubit systems, requiring only the ability to perform controlled versions of the $\SWAP$ operation between pairs of target and instruction qubits~\cite{gao_entanglement_2019,pichler_measurement_2016,marvian_universal_2016,kimmel_hamiltonian_2017}.
% The use of quantum instructions implemented with density matrix exponentiation has broad applications in quantum state metrology~\cite{lloyd_quantum_2014}, guaranteed private quantum software~\cite{marvian_universal_2016}, efficient measurement of large-scale entangled quantum systems~\cite{pichler_measurement_2016} and quantum machine learning~\cite{Biamonte2017}.

\section*{End notes}
% Acknos
\textbf{Acknowledgements:} The authors acknowledge feedback on the manuscript from Antti Veps\"al\"ainen, Cyrus Hirjibehedin, Kyle Serniak, Thomas Hazard, and Steven Weber.
MK acknowledges support from the Carlsberg Foundation during part of this work.
AG acknowledges funding from the 2019 Google US/Canada PhD Fellowship in Quantum Computing.
IM acknowledges funding from NSF grant FET-1910859.
This research was funded in part by the U.S. Army Research Office Grant W911NF-18-1-0411 and the Assistant Secretary of Defense for Research \& Engineering under Air Force Contract No. FA8721-05-C-0002.
Opinions, interpretations, conclusions, and recommendations are those of the authors and are not necessarily endorsed by the United States Government.
\\
\textbf{Author contributions:} MK, MS, AG, GS, AB, CM, and PK performed the experiments. MK, MS, AG, GS, MO, CM, and KO developed analytical tools and analyzed data. MK, MS, AG, AB, JB, RW, YS, and SG developed experimental control tools. MK, MO, CM, and KO performed algorithm simulations under realistic noise models. MK, MS, AG, GS, MO, MM, DR, SL, and IM developed the \DMEsqm{} protocol. DK, AM, BN, and JY fabricated devices.  TO, SG, and WO provided experimental oversight and support. All authors contributed to the discussion of the results and the manuscript. \\

% -------------------------------------------------------------
%  BIBLIOGRAPHY
% -------------------------------------------------------------

\end{document}

\title{Supplementary material for ``\thetitle''}

% Authors:
\author{M.~Kjaergaard$^\dagger$}\email{mortenk@mit.edu}\affiliation{\RLEaffil}
\author{M.~E.~Schwartz}\thanks{Authors contributed equally to this work}\affiliation{\LLaffil}
\author{A.~Greene}\affiliation{\RLEaffil}
\author{G.~O.~Samach}\affiliation{\RLEaffil}\affiliation{\LLaffil}
\author{A.~Bengtsson}\affiliation{\RLEaffil}\affiliation{\CHALMERSaffil}
\author{M.~O'Keeffe}\affiliation{\LLaffil}
\author{C.~M.~McNally}\affiliation{\RLEaffil}
\author{J.~Braum\"uller}\affiliation{\RLEaffil}
\author{D.~K.~Kim}\affiliation{\LLaffil}
\author{P.~Krantz}\thanks{Current address: Microtechnology and Nanoscience, Chalmers University of Technology}\affiliation{\RLEaffil}
\author{M.~Marvian}\affiliation{\RLEaffil}\affiliation{\MITMECHaffil}
\author{A.~Melville}\affiliation{\LLaffil}
\author{B.~M.~Niedzielski}\affiliation{\LLaffil}
\author{Y.~Sung}\affiliation{\RLEaffil}
\author{R.~Winik}\affiliation{\RLEaffil}
\author{J.~Yoder}\affiliation{\LLaffil}
\author{D.~Rosenberg}\affiliation{\LLaffil}
\author{K.~Obenland}\affiliation{\LLaffil}
\author{S.~Lloyd}\affiliation{\RLEaffil}\affiliation{\MITMECHaffil}
\author{T.~P.~Orlando}\affiliation{\RLEaffil}
\author{I.~Marvian}\affiliation{\DUKEaffil}
\author{S.~Gustavsson}\affiliation{\RLEaffil}
\author{W.~D.~Oliver}\affiliation{\RLEaffil}\affiliation{\LLaffil}\affiliation{\MITPHYSaffil}\affiliation{\EECSaffil}

% Make the title + abstract:
\date{\today}
\maketitle
\appendix
\onecolumngrid

% Fixing numbering of equations/figures:
% Reset counter
\setcounter{equation}{0}
\setcounter{figure}{0}
\setcounter{table}{0}
\setcounter{page}{1}

% Prefix 'S' to distinguish from equations in main matter
% \renewcommand{\theequation}{S\arabic{equation}}
\renewcommand{\thefigure}{S\arabic{figure}}
\renewcommand{\thetable}{S\arabic{table}}
\renewcommand{\bibnumfmt}[1]{[S#1]}
\renewcommand{\citenumfont}[1]{S#1}
\setcounter{figure}{0}
\setcounter{table}{0}
\renewcommand{\figurename}{EXTENDED DATA FIG.}
\renewcommand{\tablename}{EXTENDED DATA TABLE}

\tableofcontents

\section{Device parameters}
\begin{figure*}[!ht]
\center
	\includegraphics[width=1\columnwidth]{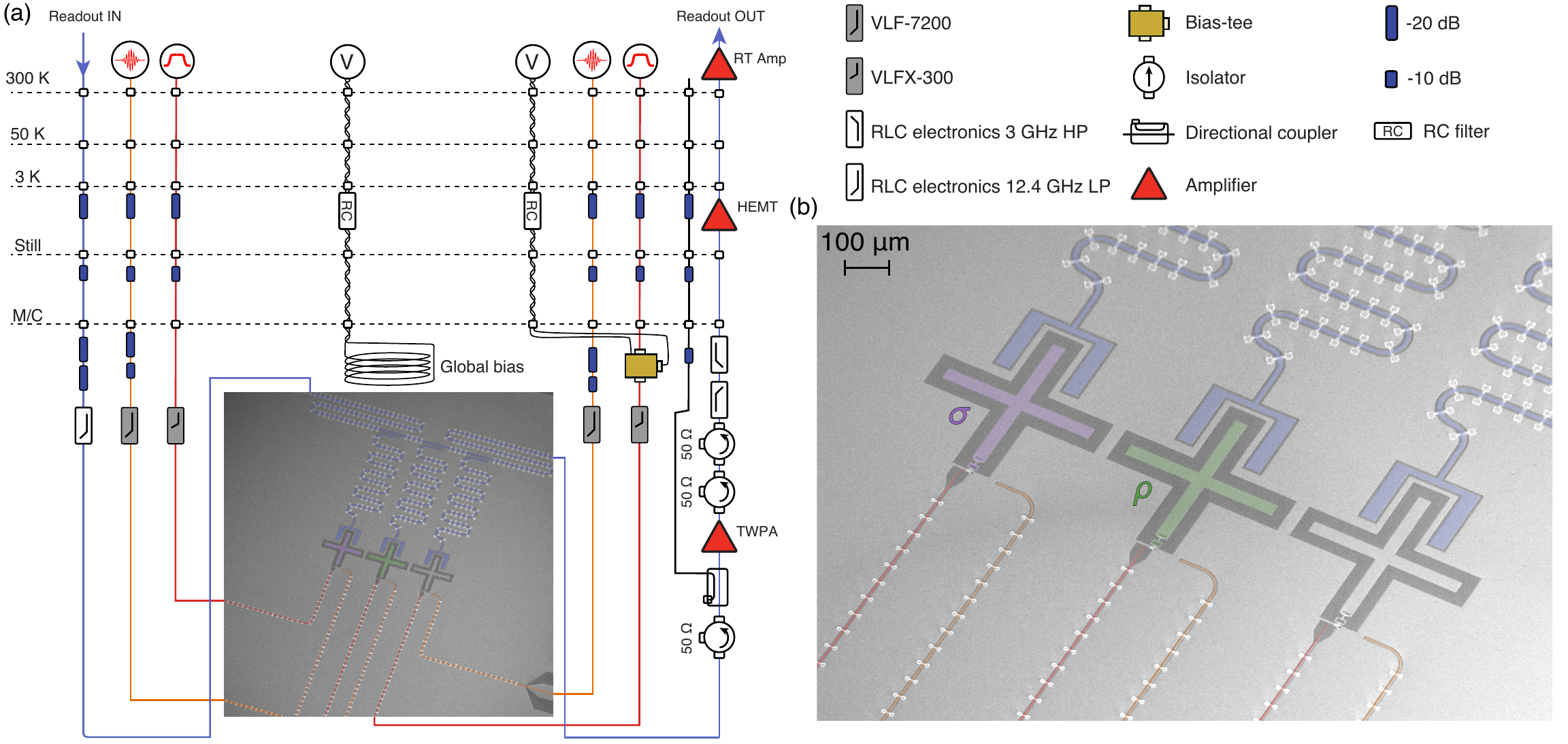}
		\caption{\footnotesize{(a) Schematic of readout- and control-wiring used for these experiments. The microwave line of qubit 3 is used to drive single-qubit gates on qubit 2. (b) SEM picture of identically fabricated device to the processor used in this work.
		}}
	\label{fig-sup:device_sem}
\end{figure*}

The quantum processor used in this work has three asymmetric `xmon'-style qubits in a linear chain~\cite{koch_charge_2007,hutchings_tunable_2017,barends_coherent_2013}.
We use the two leftmost qubits in this protocol; the third is detuned and idles in its ground state.
Extended Data Fig.~\ref{fig-sup:device_sem}a shows a schematic of the readout- and control- setup used to control the qubits.
Extended Data Fig.~\ref{fig-sup:device_sem}b shows a scanning electron micrograph of a device identical to the one used in this work.
In Extended Data Table~\ref{tbl:device_params} we summarize the parameters of the two qubits used for the experiments in the main text.
The measured lifetime $T_1$ and Ramsey coherence time $T_{2\text{R}}$ exhibit temporal fluctuations, consistent with other reports~\cite{klimov_fluctuations_2018,burnett_decoherence_2019}.

\begin{table*}[!ht]
\centering
\begin{tabular}{|c|c|c|}
\hline
                & Qubit 1               & Qubit 2            \\
Parameter       & ($\sigma$, target)    & ($\rho$, instruction)   \\
\hline
\hline
Idling frequency, $\omega_i/2\pi$  & 4.748 GHz&4.225 GHz\\
Anharmonicity, $\eta/2\pi$ & $-175$ MHz & $-190$ MHz\\
Coupling strength, $g/2\pi$ & \multicolumn{2}{c|}{10.6 MHz\quad\quad\quad}  \\
Readout resonator frequency, $f_i/2\pi$ &7.251 GHz& 7.285 GHz\\
Junction asymmetry & 1:5 & 1:10\\
\hline
\hline
Relaxation time at idling point, $T_1$ &  23 $\mu$s&  39 $\mu$s\\
Coherence time at idling point, $T_{2\text{R}}$  &  13 $\mu$s&  25 $\mu$s\\

Effective relaxation time undergoing \CZ{} trajectory, $\widetilde T_{1}$ & $\approx 17 $ $\mu$s & (same as idling) \\
Effective coherence time undergoing \CZ{} trajectory, $\widetilde T_{2\text{R}}$ & $\approx 5$ $\mu$s & (same as idling)  \\
\hline
\hline
Single-qubit gate time, $t_\text{1qb}$& 30 ns & 30 ns \\
Two-qubit gate time, $t_\text{\CZ}$  & \multicolumn{2}{c|}{60 ns\,\,\qquad\qquad} \\
\hline
\end{tabular}
\caption{Parameters of the two qubits used in this work. See text for details of the definition of $\widetilde T_1$ and $\widetilde T_{2\text{R}}$.}
\label{tbl:device_params}
\end{table*}

For a qubit undergoing frequency modulation (\textit{e.g.} to implement the \CZ{} gate), frequency-dependent $T_1$ (and $T_{2\text{R}}$) variations mean that the static coherence times do not necessarily set the relevant limiting time-scale for the qubits~\cite{klimov_fluctuations_2018}.
To account for the frequency-dependent variations in coherence as the target qubit undergoes the \CZ{} trajectory, we employ an \emph{effective} $T_1$ ($T_{2\text{R}}$) parameter, denoted $\widetilde T_1$ ($\widetilde T_{2\text{R}}$).
These effective coherence times take into account any frequency-dependent variations of coherence as the qubit frequency undergoes the trajectory to enact a $\CZ$ gate.
The effective coherence times are used in simulations of the device performance during two-qubit gates.
Since the frequency of qubit 2 is fixed during the \CZ{} gate, its effective coherence times are identical to the idling coherence times.

\begin{figure*}[!ht]
\center
    \includegraphics[width=1\columnwidth]{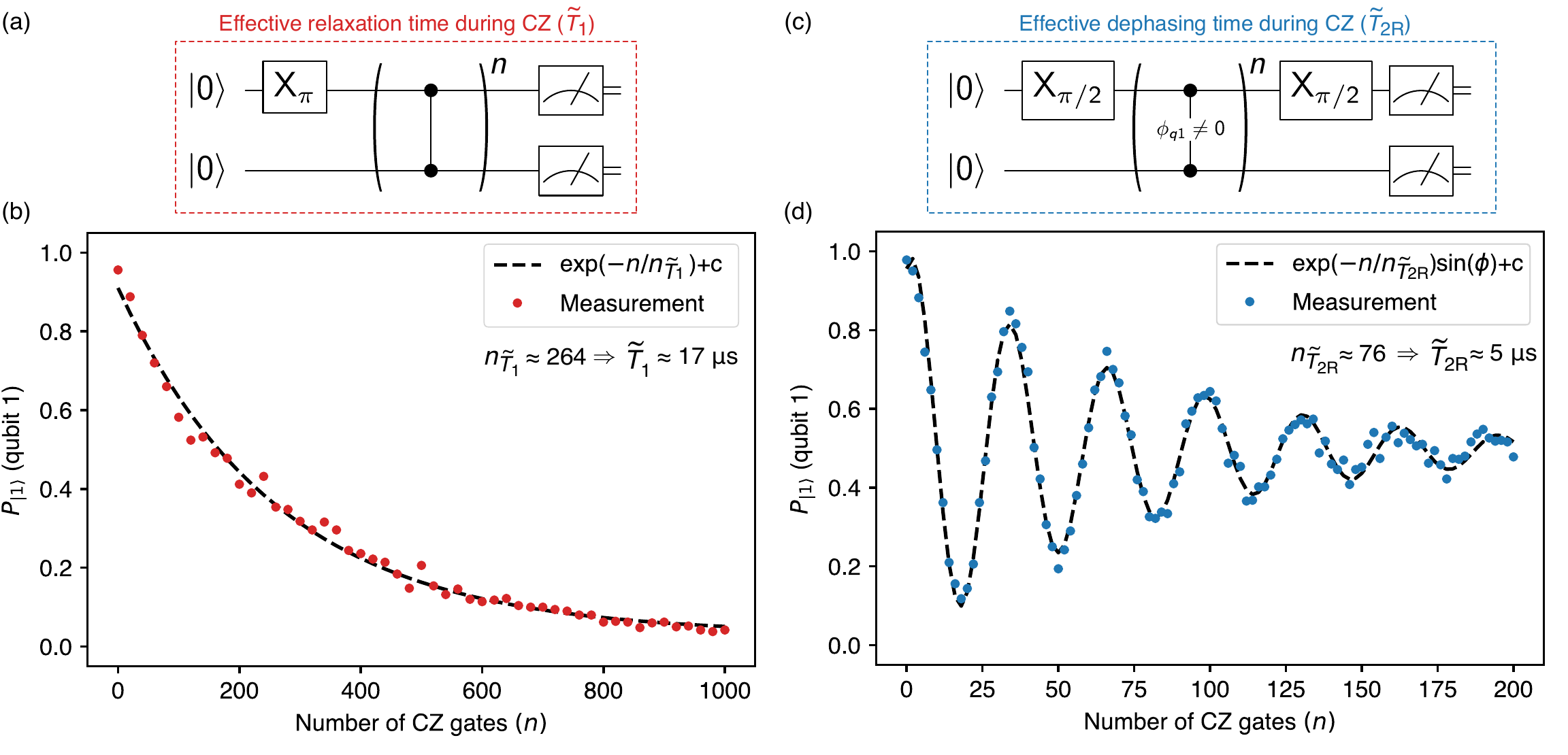}
        \caption{\footnotesize{
        (a) Measurement circuit to extract effective $T_1$-like decay time, denoted $\widetilde T_{1}$.
        (b) Probability of measuring qubit 1 in the excited state, as the number of CZ gates is increased. The number $n_{\widetilde T_1}$ sets a characteristic gate number, which can be converted into a characteristic time, $\widetilde T_1$.
        (c) Measurement circuit to extract effective $T_{2\text{R}}$-like decay time, denoted $\widetilde T_{2\text{R}}$. We essentially perform a ramsey measurement, but interleave $\CZ$ gates.
        (d) Probability of measuring qubit 1 in the excited state, as the number of CZ gates is increased. The number $n_{\widetilde T_{2\text{R}}}$ gives the effective coherence time $\widetilde T_{2\text{R}}  \approx 5 \mu $s.}}
    \label{fig-sup:effective_coherence_time}
\end{figure*}
Extended Data Fig.~\ref{fig-sup:effective_coherence_time}a shows an example measurement of $\widetilde T_1$.
We prepare the state $\ket{10}$ (an eigenstate of \CZ{}), apply $n$ \CZ{} gates in sequence, and measure the probability of staying in the $\ket{10}$ state.
The exponential decay is fitted and we find a characteristic number of gates, $n_{\widetilde T_1} \approx 264$.
The \CZ{} gate-time is 60~ns, and we use a 5~ns spacing between each pulse, leading to an effective decay time $\widetilde T_1 = n_{\widetilde T_1} \cdot t_\text{CZ} \approx 17~\mu$s.

To measure the effective coherence time $\widetilde T_{2\text{R}}$ (Extended Data Fig.~\ref{fig-sup:effective_coherence_time}b), we prepare the $\ket{+\,0}$ state, apply $n$ \CZ{} gates, and apply a final $\textsf{X}_{\pi/2}$ pulse before measuring. Unlike a standard Ramsey measurement, in which we would idle between the $\textsf{X}_{\pi/2}$ pulses, here we perform back-to-back \CZ{} gates, effectively aggregating decoherence effects over the full frequency range of the \CZ{} gate.
To ensure an oscillatory behavior, a small single-qubit phase error is added ($\phi_\text{q1} \neq 0$), equivalent to performing a detuned Ramsey experiment
Fitting an exponentially damped sine function gives a characteristic decay number $n_{\widetilde T_{2\text{R}}} \approx 76$ \CZ{} gates.
We again estimate the effective coherence time as $\widetilde T_{2\text{R}} = n_{T_{2\text{R}}} \cdot t_\text{\CZ} \approx 5~\mu$s.
\\
\\
\noindent
\section{Gate characterization}
The native gate set of our processor comprises microwave-driven single-qubit $x$- and $y$- rotations $\RX{\phi}$ and $\RY{\phi}$, single-qubit virtual-$z$ rotations $\RZ{\phi}$, and the two-qubit controlled-phase (\CPHASE{}) gate~\cite{krantz_quantum_2019}.
In particular, we calibrate a numerically optimized 99.7\% fidelity \CPHASE{} gate~\cite{kelly_optimal_2014,strauch_quantum_2003}, using the symmetrized `NetZero' optimal control waveform that reduces leakage and noise-sensitivity~\cite{rol_fast_2019,martinis_fast_2014,Oliver2005}.

We use a combination of metrics to quantify the quality of qubit operations during the algorithm.
These techniques include single- and two-qubit randomized benchmarking (RB) as well as novel techniques for amplifying and correcting coherent errors.
\begin{figure*}[!ht]
\center
	\includegraphics[width=1\columnwidth]{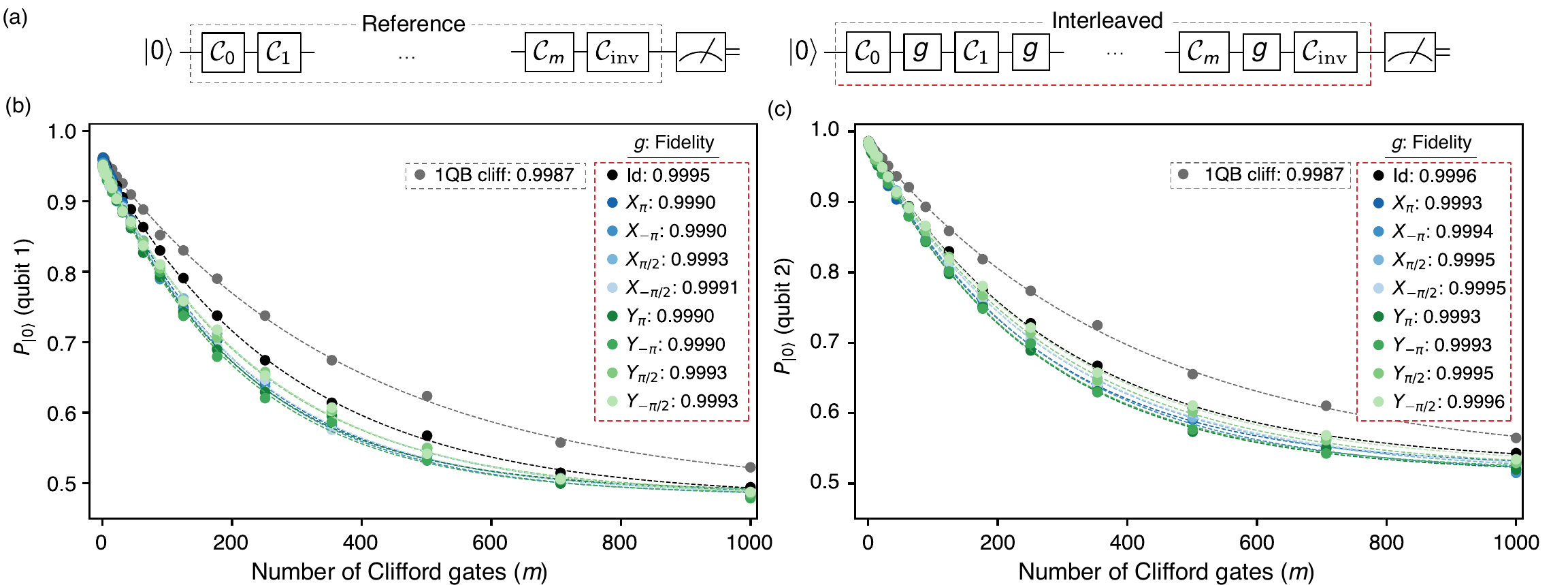}
		\caption{\footnotesize{
		(a) Circuit diagrams for measuring the reference curve (gray dashed box) and interleaved curve for a single qubit gate $g$ (red dashed box) relevant for Clifford randomized benchmarking for a single qubit.
		(b)[(c)] Results for reference (gray) and interleaved (varying colors, for each gate) randomized benchmarking for qubit 1 [qubit 2].
		}}
	\label{fig-sup:1_qubit_RB}
\end{figure*}
\begin{figure*}[!ht]
\center
	\includegraphics[width=1\columnwidth]{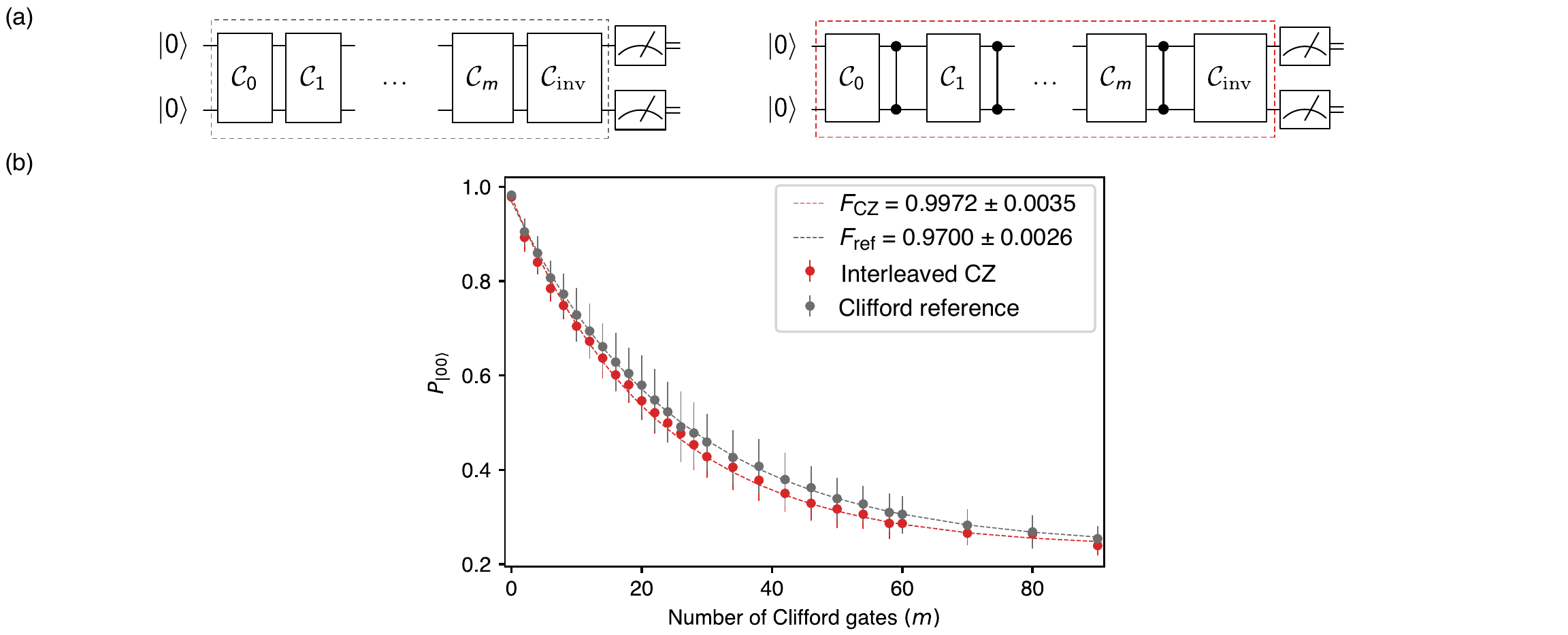}
		\caption{\footnotesize{
		(a) Gate sequences for measuring the two-qubit Clifford reference (gray dashed box) and interleaved \CZ{} (red dashed box) RB numbers.
		(b) Example decay curve of $P_{\ket{00}}$ as the number of two-qubit Clifford gates ($m$) is increased. Each datapoint is averaged over $k = 48$ randomizations of the choice of Clifford gates. Error bars are $1\sigma$ standard deviations at each point from the 48 measurements, and fitting is performed using forward propagation of points weighted by their error bars.
		}}
	\label{fig-sup:two_qubit_RB}
\end{figure*}
Extended Data Fig.~\ref{fig-sup:1_qubit_RB} shows single-qubit Clifford randomized benchmarking of the single-qubit operations on both qubit 1 (panel b) and 2 (panel c).
Each trace averages 25 randomizations of the RB circuit~\cite{barends_superconducting_2014}.
The reference curves (circuit diagram in panel a, grey dashed box) are fit to a function of the form
\begin{equation}
f(m) = Ap^m + B
\label{eq:reference_fit}
\end{equation}
For the one qubit Clifford reference curve we denote $p$ by $p_\text{r}$. The average error per Clifford gate $\mathcal{C}$ can be calculated as
\begin{equation}
\epsilon_\text{r} = \frac{1}{2}(1-p_\text{r})
\label{eq:1qb_error_eq}
\end{equation}

The error associated with a specific single-qubit gate is extracted by performing interleaved randomized benchmarking (IRB).
We fit the IRB data (circuit diagram in panel a, red dashed box) for the relevant gate (denoted $g$) to Eq.~\eqref{eq:reference_fit} (denoting by $p_g$ the $p$ value for gate $g$). Then normalizing the error rate to the one qubit Clifford reference~\cite{chen_metrology_2018},
\begin{equation}
\epsilon_{g} = \frac{1}{2}(1-p_g/p_\text{r}),
\end{equation}
Using this procedure we find an average Clifford gate fidelity ($F_\text{r} = 1-\epsilon_\text{r}$) of 0.9987 for qubit 1 and 0.9987 for qubit 2.
The average gate fidelity (\textit{i.e.} $\bar F = \langle1-\epsilon\rangle_g$) over all single-qubit gates is 0.9991 for qubit 1 and 0.9994 for qubit 2.

In Extended Data Fig.~\ref{fig-sup:two_qubit_RB}, we assess the two-qubit gate fidelity using randomized benchmarking.
The protocol is identical to the single-qubit case, except we measure the probability of being in the $\ket{00}$ state after the sequence~\cite{barends_superconducting_2014}.
We use $48$ randomizations for both reference and interleaved measurements (circuits shown in panel a).
In panel b we show the result of the RB and IRB measurements.
The error bars are $1\sigma$ standard deviations of the output distribution of the $48$ random circuits.
The fit is again performed using Eq.~\eqref{eq:reference_fit}, and error margins are extracted using forward-propagation of weights based on the standard deviation at each $m$ to ensure accurate error bounds.
This is achieved using the \texttt{absolute\_sigma} option of the Python \texttt{scipy.optimize.curve\_fit} function.
The two-qubit Clifford reference error rate is calculated similarly to Eq.~\eqref{eq:1qb_error_eq} (with $p$ being the two-qubit Clifford reference value, denoted $p_\text{2r}$) but the error per Clifford is modified to
\begin{equation}
\epsilon_\text{2r} = \frac{3}{4}(1-p_\text{2r}).
\end{equation}
Then, $\epsilon_\CZ$ is found by performing IRB and fitting the interleaved data to get $p_\CZ$ and normalizing to the 2QB reference error.
Doing so, we find a \CZ{} gate fidelity
\begin{equation}
F_\CZ = 1 - \epsilon_{\CZ} = 0.9972 \pm 0.0035.
\end{equation}
To achieve `last-mile' improvements in fidelity we use numerical optimization techniques to fine-tune parameters of the NetZero waveform, with the RB decay curve as a cost function~\cite{kelly_optimal_2014,rol_fast_2019}.
\\
\\
\noindent
\section{Coherent error reduction}
As practitioners of quantum computing have explored more complex circuits at greater depth and with more underlying structure, it has become evident that RB is a limited metric for the performance of a gate (see \textit{e.g.}~\cite{wallman_estimating_2015,proctor_what_2017,wallman_randomized_2018} and references therein).
In particular, small coherent errors can cause disproportionately deleterious effects in algorithms with a repetitive structure (such as Trotterized algorithms), and RB is ill-suited to characterize such small coherent errors because it is designed to randomize over them.

\begin{figure*}[!ht]
\center
	\includegraphics[width=1\columnwidth]{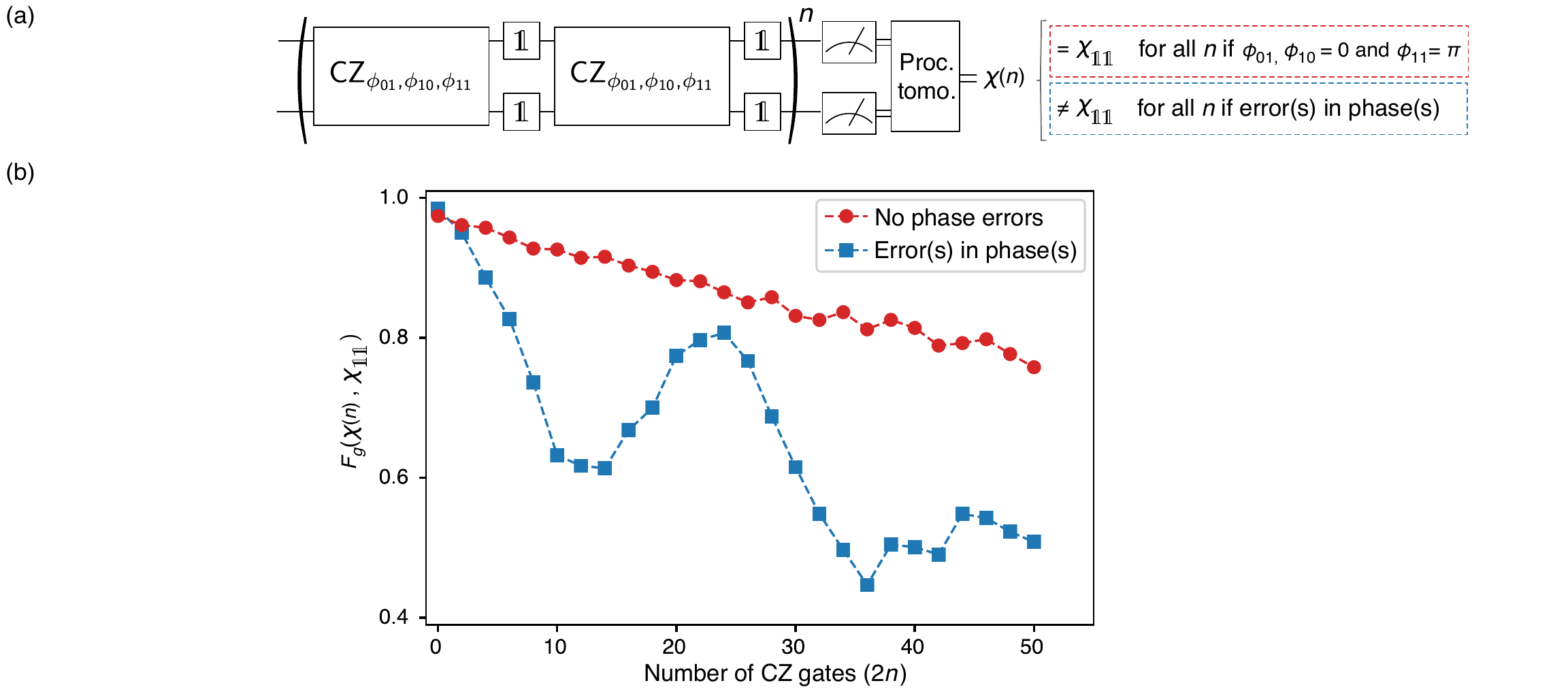}
		\caption{\footnotesize{
		(a) Gate sequence used to perform process tomography of a sequence of an even number of \CZ{} gates, to get the chi-matrix $\chi(n)$, used to compare with the identity process map to infer coherent errors.
		The gate-sequence will nominally implement $\chi_{\Id \Id}$ up to overall system decoherence (visible as the overall decrease of both the linear and oscillating measurements) if there are no phase errors in the $\CZ_{\phi_{01},\phi_{10},\phi{11}}$ gate.
		(b) The gate fidelity $F_g(\chi(n), \chi_{\Id \Id})$ as the number of \CZ{} gates ($2n$) is increased. With no phase errors in the \CZ{} gate, $F_g$ decreases monotonically. With a phase error in the \CZ{} gate $F_g$ will oscillate, with the period indicating the scale of the phase error.
		}}
	\label{fig-sup:nCZ_processtomo}
\end{figure*}

To minimize the effects of coherent errors in the \CZ{} gate, we implement a calibration technique which relies on process tomography of long strings of \CZ{} gates (Extended Data Fig.~\ref{fig-sup:nCZ_processtomo}).
The general controlled-phase gate (denoted $\textsf{CZ}_{\phi_{01}, \phi_{10}, \phi_{11}}$) is given by
\begin{equation}
\textsf{CZ}_{\phi_{01}, \phi_{10}, \phi_{11}} = \begin{bmatrix}
1 & 0 & 0 & 0 \\
0 & e^{-i\phi_{01}} & 0 & 0 \\
0 & 0 & e^{-i\phi_{10}} & 0 \\
0 & 0 & 0 & e^{-i\phi_{11}}
\end{bmatrix}
\end{equation}
If $\phi_{01} = \phi_{10} = 0$ and $\phi_{11} = \pi$ this produces the target \CZ{} gate.
However, for small deviations from these parameters it is still possible to achieve $\gtrsim 0.99$ randomized benchmarking fidelities.
Since small phase deviations can compound to form larger errors -- specifically in algorithms with a repeating pattern like \DME{} or quantum error correction protocols -- we have developed other calibration strategies to detect and correct such errors.

Our amplification protocol is comprised of implementing a circuit with two back-to-back blocks of $\CZ_{\phi_{01}, \phi_{10}, \phi_{11}}$ followed by identity gates on both qubits designed to mimic the presence of single-qubit gates, as shown in Extended Data Fig.~\ref{fig-sup:nCZ_processtomo}a.
If the \CZ{} gate contains no phase errors, this sequence produces an identity operation, irrespective of the number ($n$) of such two-\CZ{} blocks applied.
We perform two-qubit process tomography to extract the process matrix $\chi(n)$.
We compare $\chi(n)$ to the process map of a two-qubit identity operation ($\chi_{\Id \Id}$) via the gate fidelity $F_g(\chi(n), \chi_{\Id \Id})$ which is related to the process fidelity (defined in the Methods section ``State and process tomography'') according to
\begin{equation}
F_g(\chi, \chi') = \frac{dF_p(\chi,\chi') +1}{d+1}
\end{equation}
where $d$ is the dimensionality of the Hilbert space ($d=4$ in the case of a two-qubit gate).

Extended Data Fig.~\ref{fig-sup:nCZ_processtomo}b shows the gate fidelity of a circuit optimized to remove phase errors from the \CZ{} gate (red circles), and one in which a \CZ{}-gate \emph{with} phase errors is used (blue squares).
In the optimized case, the monotonic gate fidelity decay stems only from decoherence effects.
However, in the presence of a coherent phase error, the gate fidelity oscillates with $n$.
In this specific example, after roughly 25 \CZ{} gates, the phase-error has effectively rotated by $2\pi$, corresponding to an approximate per-step error of $2\pi/25 \approx 0.08\pi$ in one of the phases.
The evolution of the process maps is useful both practically (for achieving higher performance gates) and scientifically (for understanding the limitations of RB).
By examining the details of the process maps, we are able to infer in which of the parameters $\phi_{01}$, $\phi_{10}$ or $\phi_{11}$ the error appeared, and to correct accordingly.
This minor correction typically does not change the fidelity as measured with RB (except in the case of particularly egregious phase errors).
From Extended Data Fig.~\ref{fig-sup:nCZ_processtomo}b it is also clear that process tomography of a single \CZ{} instance does not reveal the coherent error: the first datapoint for the sequence with phase errors has nearly identical fidelity to the optimized gate.
Both of these facts are consistent with a growing understanding that RB may not be the optimal approach to identifying and correcting coherent errors in single- and multi-qubit gates.
Finally, the identity gates are inserted between the \CZ{} gates to as closely as possible mimic the generic optimal gate-sequence of a two-qubit algorithm, without exploiting any specific structure of an algorithm.
\\
\\
\noindent
\section{Compilation}
We implement $\delta\SWAP{}$ using single-qubit gates and the entangling $\CPHASE{}$ gate.
$\delta\SWAP{}$ has an optimal decomposition~\cite{vatan_optimal_2004}
\begin{equation}
\delta \SWAP =
 \raisebox{0.45cm}{
	\Qcircuit @C=0.5em @R=1.em{
	& \gate{\ghostgatesize} &  \ctrl{1}  & \gate{\ghostgatesize} &  \ctrl{1} & \gate{\ghostgatesize} &  \ctrl{1} & \gate{\ghostgatesize} & \qw  \\
	& \gate{\ghostgatesize} &  \ctrl{-1} & \gate{\ghostgatesize} &  \ctrl{-1}& \gate{\ghostgatesize} &  \ctrl{-1}& \gate{\ghostgatesize} & \qw
	}
 } :=
  \raisebox{0.45cm}{
  \Qcircuit @C=0.5em @R=2em{
	& \qswap 	  & \qw \\
	& \qswap \qwx & \deltaGateWhite \qw&
	}
 }
\label{eq:dSWAP_circuit}
\end{equation}
where each $\Qcircuit @C=0.5em @R=1.4em{ & \gate{\ghostgatesize} & \qw}$ represents a general single-qubit gate that depends on the value of $\delta$ and \raisebox{0.175cm}{$\Qcircuit @C=0.5em @R=0.75em{ & \ctrl{1} & \qw \\ & \ctrl{-1} & \qw}$} is the \CPHASE{} gate.
The open-source software package \texttt{Cirq}~\cite{cirq-v0.5.0_python_nodate} is used to determine the appropriate single-qubit gate parameters for a given $\delta\SWAP$.
Our $\delta\SWAP$ construction allows us to rely solely on high-fidelity gates whose performance can be validated and efficiently optimized.

A conceptually transparent approach to generating a $\delta\SWAP$ uses the decomposition
\begin{equation}
\delta \SWAP :=
  \raisebox{0.45cm}{
	\Qcircuit @C=0.5em @R=1.em{
	& \qw &  \ctrl{1}  & \gate{\textsf{H}} 	&  \ctrl{1} & \gate{\textsf{H}} 	  & \ctrl{1} &  \qw  \\
	& \qw &  \targ 	   &\qw 				&  \ctrl{-1} & \deltaGateWhiteLargeFactor \qw & \targ &  \qw
	}
 }
 \end{equation}
where
\begin{equation}
  \raisebox{0.45cm}{
	\Qcircuit @C=0.5em @R=1.75em{
	 	&  \ctrl{1} & \qw 	 \\
	   &  \ctrl{-1} & \deltaGateWhiteAgain \qw
	}
 }
 =
 \begin{pmatrix}
		1 & 0 & 0 & 0 \\
		0 & 1 & 0 & 0 \\
		0 & 0 & 1 & 0 \\
		0 & 0 & 0 & e^{-i\delta}
	\end{pmatrix}
	:=
 \textsf{CZ}_\delta
\end{equation}
is a partial $\CZ{}$ gate and \raisebox{0.175cm}{$\Qcircuit @C=0.5em @R=0.5em{ & \ctrl{1} & \qw \\ & \targ & \qw}$} is the \textsf{CNOT} gate with qubit 2 as the target.
The $\CZ_\delta$ gate can in turn be compiled using an additional decomposition
\begin{equation}
  \raisebox{0.45cm}{
	\Qcircuit @C=0.5em @R=1.9em{
	 	&  \ctrl{1} & \qw 	 \\
	   &  \ctrl{-1} & \deltaGateWhiteAgain \qw
	}
 }
 =
   \raisebox{0.45cm}{
	\Qcircuit @C=0.5em @R=0.5em{
	 	& \gate{\textsf{Z}_{\delta/2}}  &  \ctrl{1} & \qw 	 			& \ctrl{1} & \qw\\
	    & \gate{\textsf{Z}_{\delta/2}} &  \targ & \gate{\textsf{Z}_{-\delta/2}} & \targ &\qw
	}
 }.
\end{equation}
However, such an approach would introduce two \CZ{} gates for each $\CZ_\delta$ gate, adding significant circuit depth overhead.
\begin{figure*}[!t]
\center
	\includegraphics[width=1\columnwidth]{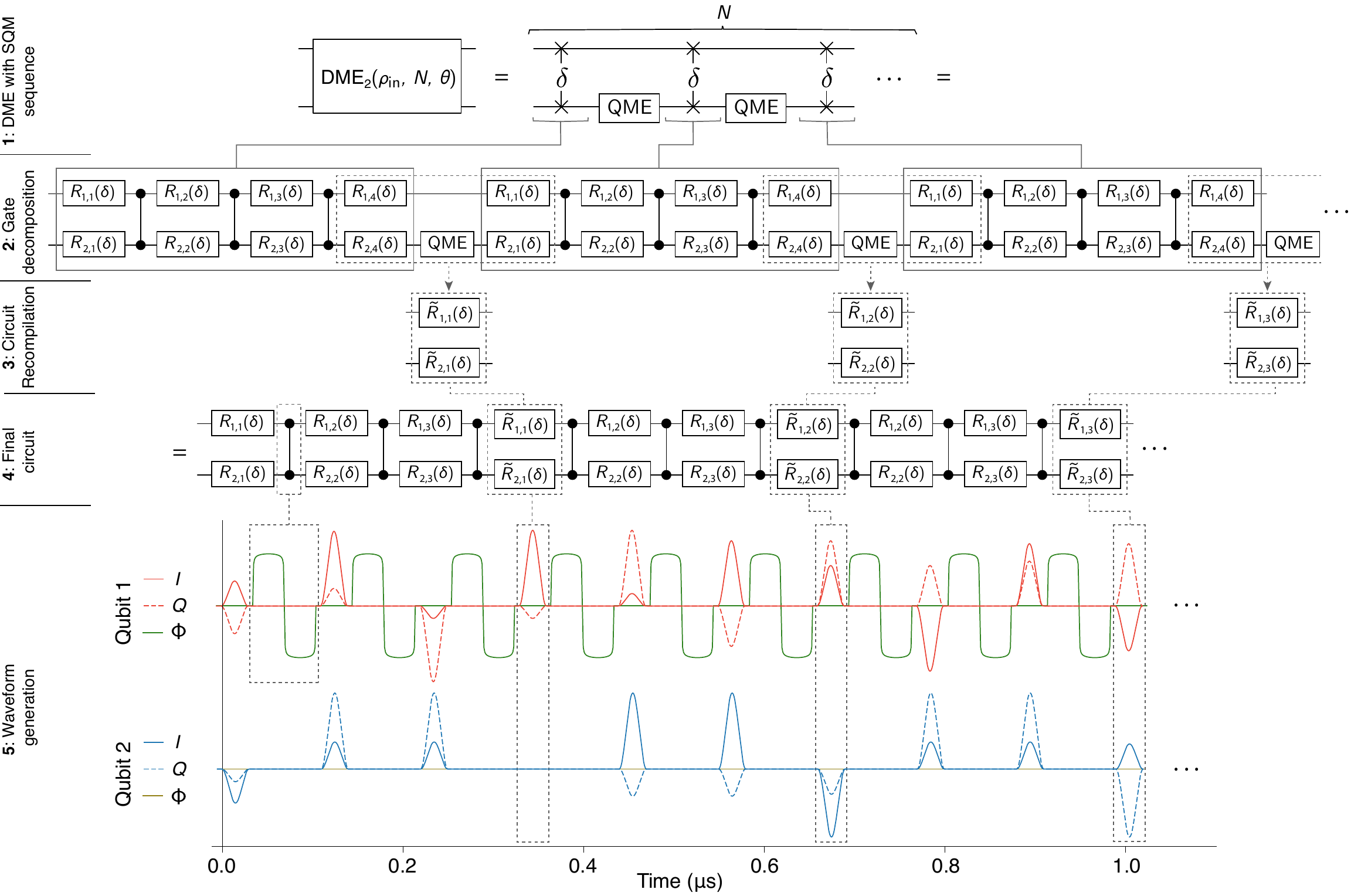}
		\caption{\footnotesize{Details of $\delta\SWAP$ and \DME{} compilation.
		\textbf{Row 1}. The density matrix exponentiation algorithm implemented using partial \SWAP{} operations and the simulated quantum measurement (\SQM{}) gate.
		\textbf{Row 2}. Decomposing each $\delta \SWAP$ according to Eq.~\eqref{eq:arbitrary_2qb_circuit}. Each substep at this step requires 8 layers of gates (7 for $\delta \SWAP$ decomposition and 1 for \SQM{}).
		\textbf{Row 3}. The three layers of single-qubit gates stemming from the the end of the $\delta \SWAP$ of step $n$, followed by \SQM{}, and the first layer of single-qubit gates in $\delta \SWAP$ of step $n+1$ can be recompiled into a single layer.
		\textbf{Row 4}. The recompiled gates are reinserted into the algorithm result in the optimal structure of exactly one \CZ{} gate, followed by a single layer of single-qubit gates.
		\textbf{Row 5}. Example waveform output to the $I, Q$ ($x,y$) ports and the flux tuning pulse (labeled $\Phi$) implementing the `NetZero' waveform used to implement the \CZ{} gate~\cite{martinis_fast_2014,rol_fast_2019}.
		}}
	\label{fig-sup:compilation}
\end{figure*}
We use a more generalized and gate-efficient approach, relying on the fact that any two-qubit gate can generically be decomposed into a circuit with the structure~\cite{vatan_optimal_2004,nielsen_quantum_2011}
\begin{equation}
U_\text{2QB} =  \raisebox{0.55cm}{
	\Qcircuit @C=0.5em @R=1.em{
	& \gate{\textsf{R}_{1,1}} &  \ctrl{1}  & \gate{\textsf{R}_{1,2}} &  \ctrl{1} & \gate{\textsf{R}_{1,3}} &  \ctrl{1} & \gate{\textsf{R}_{1,4}} & \qw  \\
	& \gate{\textsf{R}_{2,1}} &  \targ 	  & \gate{\textsf{R}_{2,2}} &  \targ   & \gate{\textsf{R}_{2,3}} &  \targ& \gate{\textsf{R}_{2,4}} & \qw
	}
 }\,.
\label{eq:arbitrary_2qb_circuit}
\end{equation}
Here $\textsf{R}_{i,j}$ is a single-qubit gate acting on qubit $i$ at moment $j$ in the circuit.

By using the identity
\begin{equation}
\raisebox{0.35cm}{
	\Qcircuit @C=0.5em @R=1.25em{
	 &  \ctrl{1}  & \qw \\
	 &  \targ 	  & \qw
	}
 }
=
\raisebox{0.35cm}{
	\Qcircuit @C=0.5em @R=1.em{
	& \qw &  \ctrl{1}  & \qw & \qw\\
	& \gate{\textsf{H}} &  \ctrl{-1} 	  & \gate{\textsf{H}} & \qw
	}
 }
\end{equation}
and absorbing the Hadamard gates ($\textsf{H}$) into the neighboring single-qubit gates, the circuit in Eq.~\eqref{eq:arbitrary_2qb_circuit} becomes identical to the circuit in Eq.~\eqref{eq:dSWAP_circuit}.

We use the open-source software \texttt{Cirq}~\cite{cirq-v0.5.0_python_nodate} to determine the settings of the single-qubit gates for each value of $\delta$.
The single-qubit rotations around the $x,y$ axes are decomposed according to $\RZ{-\varphi}\RX{\theta} \RZ{\varphi}$ (the \texttt{PhasedXPowGate} in \texttt{Cirq}) and the \textsf{R}$_Z$ rotations are performed virtually~\cite{mckay_efficient_2017}.
The $\delta \SWAP$ is implemented using the \texttt{SwapPowGate} function in \texttt{Cirq} (the \texttt{SwapPowGate} has a factor of 2 difference, relative to our definition of $\delta \SWAP$).
Thus, we are able to compose a unique composite gate sequence for each $\delta\SWAP$ relying only on high-fidelity single- and two-qubit gates.

Extended Data Fig.~\ref{fig-sup:compilation} shows the full compilation protocol.
To construct the full $\DME(\rho, N, \theta)$ circuit, we append $N$ copies of the compiled $\delta\SWAP$ gate using $\delta = \theta/N$, interleaving the requisite $\SQM_\SQMsubscript$ on qubit 2 (the instruction qubit, $\rho$) to emulate the effect of measurements.
Rows 1 and 2 show the generic structure and gate decomposition of our implementation of \DMEsqm.
The final layer of single-qubit gates in the $\delta \SWAP$ at step $n$ can be recompiled together with the $\SQM_\SQMsubscript$ and the first layer of single-qubit gates in the $\delta \SWAP$ at step $n+1$.
We use \textsf{Cirq} to slice out these three layers (Row 2) of single-qubit gates, recompile them into a single layer (Row 3), and reinsert them (Row 4).
Finally, in Row 5 we show an example waveform output from our signal generation software, implementing the first $n = 3$ steps in a $N = 5$ \DMEsqm{} program.

Our compilation relies upon a restricted set of gates that are readily characterized and numerically optimized.
The final compiled circuit has a regular structure (each \CZ{} is followed by exactly one layer of single-qubit gates), amenable to generic tuneup protocols for reducing coherent error buildup.
These features enable it to achieve high algorithmic fidelity at significant circuit depth.
\\
\\
\noindent
\section{State and process tomography}
Quantum state tomography is performed by taking advantage of independent single-shot readout of all four computational states $ \{00, 01, 10, 11\}$.
We first calibrate the measurement operators by building a matrix $\betamat$ that maps the two-qubit Pauli matrices $\hat\sigma_{\Id\Id}, \hat\sigma_{\Id Z}, \hat\sigma_{Z \Id}$, and $\hat\sigma_{ZZ}$ onto the measurement probabilities $\text{p}_{ij}$:
\beq
\label{eq:beta}
	\va{\text{p}} = \betamat \va{\sigma},
\eeq
where
\beq
	\va{\text{p}} \equiv
	\begin{pmatrix}
		\text{p}_{00} \\ \text{p}_{01 } \\ \text{p}_{1 0} \\ \text{p}_{1 1 } \\
	\end{pmatrix}
	\qq{and}
	\va{\sigma} \equiv
	\begin{pmatrix}
		\pauli{\Id\Id} \\ \pauli{\Id Z} \\ \pauli{Z\Id} \\ \pauli{ZZ} \\
	\end{pmatrix}
\eeq
The $\betamat$ matrix is calibrated using techniques drawn from Ref.~\cite{Chow_2010}; a full motivation and derivation of the technique can be found there.
For a measurement of $\va{\text{p}}$ with perfect fidelity and no qubit decay during measurements, all components of $\betamat$ have amplitude 0.25; deviations from this amplitude correspond to a calibration of such measurement errors.
 We begin by calibrating the single-qubit $\betamat$ matrices, namely

\beq
\label{eq:}
\begin{pmatrix}
	\text{p}_{0} \\ \text{p}_{1}
\end{pmatrix}
=
\begin{pmatrix}
	\beta^0_\Id & \beta^0_\mathrm{Z} \\
	\beta^1_\Id & \beta^1_\mathrm{Z}
\end{pmatrix}
\begin{pmatrix}
	\pauli{\Id } \\ \pauli{Z}
\end{pmatrix}
\eeq
by fitting Rabi oscillations in p$_0$ and p$_1$ for each qubit. Because the two-qubit probability vector $\va{\text{p}}$ is generated from correlations between single-qubit measurements, the two-qubit $\betamat$ matrix is given by the tensor product of the single-qubit matrices, \textit{e.g.} $\betamat = \betamat_1 \otimes \betamat_2$.

An arbitrary $4\times 4$ matrix, including a two-qubit density matrix $\rho$, may be mapped onto the Pauli basis according to
\beq
	\rho = \sum_{ i,j = \{\Id,X,Y,Z\}} c_{ij}\pauli{ij}.
\eeq
The general $4\times 4$ matrix of this form has sixteen degrees of freedom; trace normalization of a physical density matrix reduces this to fifteen.
The native readout gives us access to the components of $\rho$ contained in $\hat\sigma_Z$.
We gain information about the other components by performing one of nine pre-measurement rotations drawn from:
\beq
\textsf{R} = \textsf{R}_1 \otimes \textsf{R}_2
\eeq
where
\beq
\textsf{R}_{1,2} =
\begin{cases}
\RY{-\frac{\pi}{2}} & \qq{mapping} \pauli{X} \mapsto \pauli{Z}\\
\RX{\frac{\pi}{2}} & \qq{mapping} \pauli{Y} \mapsto \pauli{Z}\\
\Id & \qq{mapping} \pauli{Z} \mapsto \pauli{Z}
\end{cases}
\eeq
For data in Figure~\figDMEbloch~(\figDMEfidelity, \figDMEprocessfidelity) we perform 2000 (500) single-shot measurements for each tomographic rotation in order to ensure accurate estimates of $\va{\text{p}}$.
Each of the nine rotation-and-measurement pairings provides four linearly independent measurements of a form similar to Eq. \eqref{eq:beta}, for a total of thirty-six equations that over-specify fifteen degrees of freedom.
We perform maximum-likelihood estimation~\cite{Banaszek_1999} to derive the positive semi-definite Hermitian matrix that is most consistent with our combined measurement results.

Single-qubit density matrices in Figures~\figDMEbloch--\figDMEfidelity~are extracted by performing partial traces over the two-qubit density matrix calculated using the approach described above; the data in Figure~\figDMEprocessfidelity~are drawn from single-qubit tomography performed on the target qubit using a similar protocol.

Single-qubit quantum process tomography, as presented in Figure~\figDMEprocessfidelity, is performed using standard techniques~\cite{nielsen_quantum_2011}. The target qubit is sequentially prepared in four input states
\beq
\label{eq:ptomo-prerots}
\targinit = \left\{ \ketbra{0}, \ketbra{1}, \ketbra{+}, \ketbra{i}\right\}
\eeq
which span the single-qubit Hilbert space.
These prepared states are then passed through the process $\DMEsqm(\ctrlinit, N, \theta)$ and single-qubit state tomography is performed to extract the set of mappings $\{\targinit~\xmapsto{\DMEsqm(\ctrlinit, N, \theta)}~\sigma(N)\}$.
Linear combinations of these mappings provide the process map $\chi$ that reveals the effect of the quantum channel on an arbitrary input density matrix.
We then employ techniques developed in Ref.~\cite{Knee_2018} to efficiently project $\chi$ onto the closest completely positive and trace-preserving (CPTP) mapping $\chi_\text{CPTP}$, ensuring physicality of the process.
\\
\\
\noindent
\section{Bootstrap error analysis}
We employ bootstrapping techniques to derive the uncertainty bounds in Figs. \figDMEfidelity--\figDMEprocessfidelity.
In principle, one could simply take a sample of many \SQM{} randomizations and calculate the mean and uncertainty within that dataset.
However, those error bars are not representative of the error in the \DMEsqm{} protocol -- rather, they represent the uncertainty of a protocol in which only a single \SQM{} randomization is used to perform \DMEsqm.
As a result, these error bars are unphysically large, particularly at small $N$ where the protocol chooses from one of only a few paths that have very different outcomes.

The true uncertainty of the \DMEsqm{} protocol is captured by $i$) accumulating enough \SQM{} samples to ensure sufficient randomizations, $ii$) building density/process matrices from the average outcome of all these randomizations, and then $iii$) repeating this process many times with different randomizations to estimate the uncertainty.
This is precisely what bootstrapping accomplishes~\cite{efron1979}.

The following describes the protocol for extracting boostrapped averages and uncertainties for Figure~\figDMEfidelity. For each data point representing a unique setting of $\DMEsqm(\ctrlinit, N, \theta)$, we employ the following protocol:

\begin{enumerate}[parsep=0pt, itemsep=0pt, label*=\arabic*.]
\item{For a given instantiation of the \SQM{} gates, execute $\DMEsqm(\ctrlinit, N, \theta)$ and perform two-qubit state tomography.}
\item{For $r_\SQM$ different instantiations of \SQM{} gates, repeat step 1 to accumulate the experimental density matrices from which bootstrapped samples will be drawn.}
\item{Using sample-with-replacement, select $n_\textrm{samp}$ samples from the $r_\SQM$ datasets and average the density matrices together. This represents a single bootstrapped density matrix.}
\item{Perform a partial trace over the instruction qubit to extract the reduced density matrix of the target system.}
\item{Calculate the state fidelity to the states of interest.}
\item{Repeat steps 3-5 a total of $N_\textrm{samp}$ times to extract mean fidelities and $1\sigma$ uncertainties.}
\end{enumerate}

State fidelity is calculated according to \cite{wilde_2017}
\beq
F_s(\sigma, \sigma') =  \Tr\left(\sqrt{\sqrt{\sigma'} \sigma \sqrt{\sigma'}}\right)^2.
\eeq

The bootstrapping protocol for generating process maps and process fidelities in Figure~\figDMEprocessfidelity is similar to that used for state tomography, but we lay it out here explicitly for completeness.

\begin{enumerate}[parsep=0pt, itemsep=0pt, label*=\arabic*.]
\item{For a given instantiation of the \SQM{} gates, prepare the target input states $\{\targinit \}$, apply $\DMEsqm(\ctrlinit, N, \theta)$, and perform single-qubit state tomography to generate the mappings $\{\targinit \mapsto \sigma(N) \}$ required for process tomography.}
\item{For $r_\SQM$ different instantiations of \SQM{} gates, repeat step 1 to produce a set of $4 \times r_\SQM$ single-qubit density matrices.}
\item{For each of the four $\targinit$, select an independent sample-with-replacement of $n_\textrm{samp}$ $\sigma_\text{out}$ instances and average together, leaving four averaged mappings $\{\targinit \mapsto \sigma_\text{out} \}$.}
\item{Calculate the process matrix using the averaged mappings $\targinit \mapsto \sigma(N)$. This represents a single bootstrapped process matrix.}
\item{Calculate the process fidelity to the process of interest.}
\item{Repeat steps 3-5 a total of $N_\textrm{samp}$ times to extract mean fidelities and $1\sigma$ uncertainties.}
\end{enumerate}

The process fidelity between two $\chi$-matrices is given by~\cite{nielsen_simple_2002}:
\beq
F_p(\chi, \chi')~=~\Tr(\sqrt{ \sqrt{\chi'}\chi\sqrt{\chi'}})^2.
\eeq

In Figure~\figDMEfidelity~we collect $r_\SQM=295$ circuit randomizations; in Figure~\figDMEprocessfidelity~we collect $r_\SQM=105$ circuit randomizations.
In both cases we use $n_\textrm{samp}=100$ and $N_\textrm{samp}=50$. The number of \SQM{} randomizations used for process tomography was limited by experimental time, due to the significant additional experimental overhead required for process tomography in comparison to state tomography, and due to the fact that in Figure~\figDMEprocessfidelity~we characterize processes for six settings of $\rho$.
The bootstrap sample size $n_\text{samp}$ and number of bootstrap samples $N_\text{samp}$ are chosen somewhat arbitrarily, as in all bootstrapping implementations, but are designed to ensure that each bootstrapped sample approaches a central limit with respect to the underlying \SQM{} randomization.
A graphical representation of the convergence under \SQM{} randomizations is shown in Extended Data Fig.~\ref{fig-sup:sqm_randomizations}; more details are provided in the Supplementary Methods.
% ------------------------------------------------------------
%  DETAILS OF SIMULATION
% -------------------------------------------------------------
\\
\\
\noindent
\section{Circuit simulation with noise}
\begin{figure*}[!t]
\center
	\includegraphics[width=1\columnwidth]{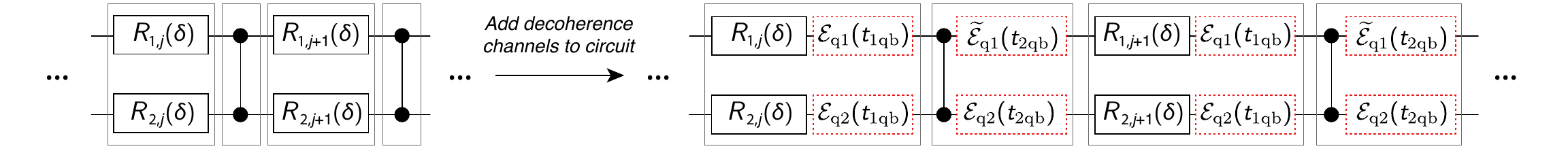}
		\caption{\footnotesize{Instrumenting the \DMEsqm{} circuit for simulation of decoherence-induced errors.
		}}
	\label{fig-sup:circuit-sim}
\end{figure*}

In order to show the qualitative consistency between the data in Figure~\figDMEfidelity~and a model of coherence-limited implementation of the $\DMEsqm{}$ protocol, we simulate the randomized \DMEsqm{} circuits with added decoherence.
We input a $\DMEsqm$ circuit generated by \textsf{Cirq} to a software tool that adds decoherence (amplitude damping and dephasing) channels corresponding to the identity for duration(s) of the preceding one- or two-qubit gate.
An example of this procedure is shown in Extended Data Fig.~\ref{fig-sup:circuit-sim}.

The channel $\mathcal E$ that composes amplitude damping and dephasing is given by
\begin{equation}
\begin{gathered}
  \mathcal{E}_{\mathrm{q}k}(t_\text{1qb}): \rho_{\text{q}k} \mapsto \\
  \sum_{\substack{i = 1, 2 \\ j = 1, 2, 3}}
   A_{i,\Gamma_1}(t_\text{1qb})  D_{j, \Gamma_\phi}(t_\text{1qb}) \rho_{\mathrm{q}k}  D^\dagger_{j, \Gamma_\phi}(t_\text{1qb})  A^\dagger_{i, \Gamma_1}(t_\text{1qb}),\\
\end{gathered}
\end{equation}
where $A_{i, \Gamma_1}(t)$ is the amplitude damping process (with $\Gamma_1~=~1/T_1$), and $D_{j, \Gamma_\phi}(t)$ is the dephasing process ($\Gamma_\phi~=~1/T_{2\text{R}}-1/2T_1$), $\Gamma_{1, \text{q}k}$ and $\Gamma_{\phi, \text{q}k}$ are the appropriate coherence parameters for qubit $k$, and $t$ is the time of the preceeding single- or two-qubit gate on that qubit. The amplitude damping and dephasing Krauss operators are given by
\begin{eqnarray}
A_{1,\Gamma_1}(t) &=& \begin{pmatrix}
  1 & 0 \\
  0 & e^{-\Gamma_{1,\mathrm{q}k} t / 2}
  \end{pmatrix},\\
 A_{2, \Gamma_1}(t) &=& \begin{pmatrix}
  0 & \sqrt{1 - e^{-\Gamma_{1,\mathrm{q}k} t}} \\
  0 & 0
  \end{pmatrix}, \label{eq-sup:amplitude_damping}
  \\
D_{1,\Gamma_\phi}(t) &=& \begin{pmatrix}
  e^{-\Gamma_{\phi,\mathrm{q}k} t / 2} & 0 \\
  0 & e^{-\Gamma_{\phi,\mathrm{q}k} t / 2}
\end{pmatrix},\\
 D_{2,\Gamma_\phi}(t)& =& \begin{pmatrix}
  \sqrt{1 - e^{-\Gamma_{\phi,\mathrm{q}k} t}} & 0 \\
  0 & 0
\end{pmatrix},\\
 D_{3,\Gamma_\phi}(t) &=& \begin{pmatrix}
  0 & 0 \\
  0 & \sqrt{1 - e^{-\Gamma_{\phi,\mathrm{q}k} t}}
\end{pmatrix}.
\end{eqnarray}
The channel $\widetilde{\mathcal {E}}$ is defined similarly to $\mathcal E$, but decoherence rates in the process definitions are replaced with their \emph{effective} coherence parameters.
The channel $\widetilde{\mathcal {E}}$ thus accounts for the modified coherence properties as qubit 1 undergoes the \CZ{} trajectory (see Extended Data Fig.~\ref{fig-sup:effective_coherence_time}).

Each instrumented circuit yields an \SQM{}-dependent density matrix representing the simulated finite-coherence circuit output for that \SQM{} realization.
These density matrices are averaged over all $2^N$ \SQM{} realizations (for a \DMEsqm{} circuit with $N$ steps), thus producing the `noisy' simulated two-qubit \DMEsqm{} output state, denoted `Sim. $F_s(\sigma, \sigma_\text{ideal})$ with decoherence' and plotted as a solid line in Figure~\figDMEfidelity b.
For the simulations presented, we used parameters $T_1 = 20~\mu$s, $T_{2\text{R}} = 10~\mu$s for both qubits, and effective coherence times for qubit 1 of $\widetilde T_1 = 10~\mu$s and $\widetilde T_{2\text{R}} = 5~\mu$s during the channel $\widetilde{\mathcal E}$. These parameters are qualitatively consistent with, but overall reduced from, the measured parameters in Table ~\ref{tbl:device_params}. This difference may indicate additional coherent errors not captured by this model (e.g. from residual $\pauli{Z}\pauli{Z}$-interaction or leakage out of the computational subspace).

% -------------------------------------------------------------
%  ALGORITHMIC ERROR IN DME
% -------------------------------------------------------------
\section{Algorithmic error in \DME}
\label{sup-sec:dme-error}
In this section we show that the algorithmic error in $\DME(\rho, N, \theta)$ (the version of DME in which the instruction state is refreshed with a new, perfect copy after each Trotter step) may be modeled as an amplitude damping channel and derive its scaling with the parameters of the algorithm. We do so first for a specific instruction state, and then generalize to an arbitrary instruction. Throughout we use $\hat\sigma_i$ to indicate the corresponding Pauli matrix.

Suppose that we have instruction and target qubits initially in states $\rho$ and $\sigma$ respectively, and apply the operation $e^{-i\SWAP\delta}$ to the joint state $\rho \otimes \sigma$.
We will first consider the special case in which $\rho = \ketbra{0}$ and then show how this generalizes to an arbitrary state.
The effect of the $\delta\SWAP$ on the target qubit is given by the quantum channel
\beq
\mathcal{E}_{\delta\SWAP}^{\rho = \ketbra{0}}(\sigma) =\Tr_\rho\Big(e^{-i \SWAP \delta} \Big[\sigma\otimes \ketbra{0} \Big] e^{i\SWAP\delta} \Big) ,
\eeq
Next, we use the fact that
\beq
e^{i\SWAP\delta}= \cos(\delta)\pauli{\Id\Id} + i\sin(\delta)\SWAP,
\eeq
which follows from the fact that $\SWAP^2=\pauli{\Id\Id}$ where $\pauli{\Id\Id}$ is the two-qubit identity matrix.
Using this together with the identity $\Tr_\rho\left(\SWAP(X\otimes Y)\right)=YX$ (where $\Tr_\rho$ is a partial trace over the second subsystem) we find
\beq
\mathcal{E}_{\delta\SWAP}^{\rho = \ketbra{0}}(\sigma) = \cos^2(\delta)\sigma+ i\cos(\delta)\sin(\delta) [\sigma,\ketbra{0}]+\sin^2(\delta) \ketbra{0}.
\eeq
Using the matrix representation of $\sigma$ in the $\{\ket{0}, \ket{1}\}$ basis, we find that $\sigma$ transforms as
\beq
\label{eq:sigma_evolution}
\begin{pmatrix}
 \sigma_{00}' & \sigma_{01}'    \\
  \sigma_{10}' &  \sigma_{11}'
\end{pmatrix}
=
\begin{pmatrix}
\sigma_{00}+\sigma_{11}\sin^2(\delta)  & \cos{\delta} e^{-i\delta}\sigma_{01}  \\
\cos{\delta} e^{+i\delta} \sigma_{10}  &  \sigma_{11} \cos^2(\delta) \end{pmatrix}
\eeq
where $\sigma_{ij}=\langle i|\sigma|j\rangle$ as measured in the $\{\ket{0}, \ket{1} \}$ basis. The channel that implements this transformation has a simple interpretation as the composition of a rotation and an amplitude decay.

Let
\beq
\mathcal{U}_{\delta}^{\rho = \ketbra{0}}(\cdot)=e^{-i\delta \ketbra{0}}(\cdot) e^{i\delta \ketbra{0}}=e^{-i\frac{\delta}{2} \pauli{Z}}(\cdot) e^{+i\frac{\delta}{2} \pauli{Z}}
\eeq
be the superoperator corresponding to the unitary $e^{-i\delta\ketbra{0}}$, or equivalently, the superoperator corresponding to the rotation by angle $\delta$ around $z$ axis. Also, let $\mathcal{A}_p$ be the amplitude damping channel described by the Kraus decomposition
\beq
\mathcal{A}_{p}(\sigma)=A_1 \sigma A^\dag_1+A_2 \sigma A^\dag_2
\eeq
where
\beq A_1=\left( \begin{array}{cc}
 1 & 0   \\
  0 &  \sqrt{1-p}
\end{array}
\right)\ \ , \  \ A_2=\left(
\begin{array}{cc}
 0 & \sqrt{p}   \\
  0 &  0
\end{array}
\right)\ .
\eeq
This amplitude damping channel describes the process in which the system in state $|1\rangle$ decays to state $|0\rangle$ with probability $p$.
 It can be shown that the amplitude damping channel satisfies the condition
 \beq\label{eq:cov}
 \mathcal{A}_p\circ \mathcal{U}_\delta = \mathcal{U}_\delta\circ \mathcal{A}_{p}
 \eeq  for all $\theta\in[0,2\pi)$.
This equality implies that the action of this channel is invariant under rotations around $z$ axis.

Then, using Eq. \eqref{eq:sigma_evolution} one can show that
\beq
\label{eq:e_dswap}
\mathcal{E}_{\delta\SWAP}^{\rho = \ketbra{0}}(\sigma)= \mathcal{A}_{\sin^2(\delta)}\circ \mathcal{U}_\delta(\sigma)=  \mathcal{U}_\delta\circ  \mathcal{A}_{\sin^2(\delta)}(\sigma)
\eeq
The overall effect of one Trotter step of $\DME_N$ can therefore be understood as the following: ($i$) Applying the  unitary $e^{-i\delta \ketbra{0}}$ to the system $\sigma$, followed by ($ii$) applying the amplitude damping channel $\mathcal{A}_{\sin^2\delta}$ to the system $\sigma$. Note that because of the condition in Eq.~\eqref{eq:cov}, by flipping the order of steps ($i$) and ($ii$) we get exactly the same final state.

Now suppose we repeat the above operation $N$ times. That is we prepare the instruction qubit in state $\rho = \ketbra{0}$, couple it to $\sigma$ via the unitary $e^{-i  \SWAP\delta}$, then discard the instruction qubit and prepare it again in state $\ketbra{0}$, and repeat the above procedure with $N$ different copies of $\rho$. Then, using Eq.~\eqref{eq:cov} one can show that, given an initial state $\sigma$, the final state of the target system will be
\beq
\left[\mathcal{E}_{\delta\SWAP}^{\rho = \ketbra{0}}\right]^N(\sigma) =
\big[ \mathcal{A}_{\sin^2(\delta)}\circ \mathcal{U}_\delta\big]^N(\sigma) =
\mathcal{A}^N_{\sin^2(\delta)}\circ \mathcal{U}_{N\delta}(\sigma).
\eeq
Since amplitude damping channels are closed under composition, we see that
\beq
\label{eq:dmeN_p_err}
\mathcal{A}^N_{\sin^2(\delta)}=\mathcal{A}_{1-\cos^{2N}(\delta)}.
\eeq
Therefore,  the overall effect on the target system is equivalent to applying the perfect unitary $e^{-i N\delta\ketbra{0}}$, and then applying the amplitude damping channel $\mathcal{A}_{1-\cos^{2N}(\delta)}$.

Now, suppose in the above procedure, instead of state $\ketbra{0}$ we prepare the instruction qubit in state $\ketbra{\phi}=V\ketbra{0}V^\dagger$, where $V$ is an arbitrary unitary.
Then, using the fact that $\SWAP(V\otimes V')=(V'\otimes V) \SWAP$, one can show that the overall effect of this transformation on the target system can be described as a unitary rotation $e^{-i N \delta \ketbra{\phi}}$ followed by an amplitude damping channel in the basis defined by state $\ket{\phi}$ and its orthogonal state.

To translate explicitly to the language of the main text, let $\delta = \theta/N$ and $\rho = \ketbra{\phi}$, and use the above procedure to implement the unitary $e^{-i\rho\theta}$ on the target system $\sigma$, using $N$ copies of the instruction state $\rho$. From Eq. \eqref{eq:dmeN_p_err} we find that the overall error in this procedure is determined by the probability $p_N=1-\cos^{2N}(\delta)$. Then, for $\delta\in (0,2\pi]$ and $N\gg 1$ we have
\beq
p_N=1-\cos^{2N}\left(\frac{\theta}{N}\right)\approx 1-e^{-\frac{\theta^2}{N}} \approx \frac{\theta^2}{N}, \quad \text{for large } N
\eeq
In the limit of large $N$, this corresponds to an algorithmic error for the $\DME_N$ algorithm of $\mathcal{O}\left(\theta^2/N\right)$, as quoted in the main text.

% -------------------------------------------------------------
%  PROOF OF THE ERROR RATE FROM USING SQM
% -------------------------------------------------------------
\section{Algorithmic error due to \SQM}\label{sup:SQM_error}
Here we provide an intuitive picture for the quantum measurement emulation (\SQM) operation as well as a formal proof of the modified algorithmic error bound in  Eq.~(\eqdmesqmcircuit) of the main paper.

We will build the intuition for this section by returning to the concrete example from Appendix \ref{sup-sec:dme-error}, \textit{i.e.} the instruction qubit prepared in $\rho = \ketbra{0}$.
We will also suppose that the target qubit is prepared in an orthogonal state, say, $\sigma = \ketbra{+i}$ (which is an eigenstate of the Pauli matrix $\pauli{Y})$.
Since $\delta\SWAP$ is a symmetric operation by the logic in Appendix \ref{sup-sec:dme-error} the state of $\rho$ following a small $\delta\SWAP$ interaction is given by a rotation about the $y$-axis followed by an amplitude damping channel (which we will neglect for the moment).
In this case, the state of the instruction qubit becomes
\beq
\rho' =
	\begin{pmatrix}
		\cos^2(\delta) & -\cos(\delta)\sin(\delta) \\
		-\cos(\delta)\sin(\delta) & \sin^2(\delta)
	\end{pmatrix}
\eeq
The trace distance between $\rho$ and $\rho'$ is of order $|\delta|$.
However, if we measure-and-forget the state of the instruction qubit in the basis of its original polarization (\textit{i.e.} the $z$-basis), the coherent off-diagonal components of the density matrix are dephased and we are left with
\beq
\rho'' =
	\begin{pmatrix}
		\cos^2(\delta) & 0 \\
		0 & \sin^2(\delta)
	\end{pmatrix}
\eeq
The trace distance between $\rho''$ and $\rho$ is of order $\delta^2$.
Because \DME{} operates in the $\delta \ll 1$ regime, we have $\delta^2 \ll \delta$.
Measuring-and-forgetting therefore leaves the instruction qubit in a slightly perturbed state that is closer to that of the initial state $\rho$.

The intuition developed for $\rho = \ketbra{0}$ extends naturally to an arbitrary initial state $\rho = \ketbra{\arbgs}$, in a basis defined by $\SQMsubscript = \left\{\ket{\arbgs},\ket{\arbes} \right\}$.
A small arbitrary rotation will result in the state
\beq
\label{eq:sqm-beta-pre}
\rho' = \cos^2(\beta)\ketbra{\arbgs} + \sin^2(\beta) \ketbra{\arbes} + \cos(\beta)\sin(\beta) \left(e^{i\phi}\ketbra{\arbgs}{\arbes} + e^{-i\phi}\ketbra{\arbes}{\arbgs} \right),
\eeq
where $\beta$ and $\phi$ generically paramterize the rotation.
A measurement in the basis $\SQMsubscript$ dephases the off-diagonal elements in this basis, leaving
\beq
\label{eq:sqm-beta-post}
\rho'' = \cos^2(\beta)\ketbra{\arbgs} + \sin^2(\beta) \ketbra{\arbes}
\eeq
which is closer than $\rho'$ to $\rho$ by a factor of $|\beta|$.

Performing a physical measurement along an arbitrary axis $\SQMsubscript$ generically would require $i$) rotating $\SQMsubscript$ onto the $z$-axis, $ii$) performing a projective readout, and $iii$) rotating back to the original axis. All of these steps require finite clock time: single-qubit gates (measurements) typically require tens (hundreds) of nanoseconds to complete.
We would like to avoid this significant experimental overhead while still maintaining the ability to partially restore the instruction qubit to its initial state.
Instead of physically performing the measurement, we can apply the unitaries $\{\pauli{\Id},\pauli{\SQMsubscript}\} $  with equal probabilities, where $\pauli{\SQMsubscript} = \hat{n}_\parallel \cdot (\pauli{X}, \pauli{Y}, \pauli{Z})$ and $\hat{n}_\parallel$ is a unit vector parallel to $\rho$.
Such protocols may be equivalently thought of as an approach to turning a coherent error into an incoherent error along a known axis.
This protocol is the quantum measurement emulation (\SQM) operation used in the main paper.

When averaged over many iterations, the randomized \SQM{} operation dephases the system in the $\SQMsubscript$ basis, just as in Eq. \eqref{eq:sqm-beta-pre}-\eqref{eq:sqm-beta-post}.
Assuming the instruction qubit is initially in state $\rho'$, it turns out that the resulting state is the same for measurement and random gate application, \textit{i.e.}
\beq
\frac{\ketbra{\arbgs} \rho' \ketbra{\arbgs}+ \ketbra{\arbes} \rho' \ketbra{\arbes}}{2}=\frac{\pauli{\Id}\rho'\pauli{\Id}+\pauli{\SQMsubscript} \rho' \pauli{\SQMsubscript}}{2}= \frac{1}{2\pi} \int_0^{2\pi} d\gamma\ e^{-i\gamma \pauli{\SQMsubscript}} \rho' e^{i\gamma \pauli{\SQMsubscript}} \ .
\eeq
These three terms represent respectively measuring-and-forgetting, random gate application, and phase randomization.
Their equivalence can be understood more formally from the standpoint of the stochastic master equation, to which Ref. \cite{Jacobs_2006} provides an accessible introduction.
This approach is also related to the Quantum Zeno Effect, in which persistent measurement along an axis of interest ``pins'' the qubit state to that axis by continuously dephasing any rotations away from it \cite{itano1990quantum}.

Finally, we calculate the additional error introduced to the \DME{} algorithm by the use of \SQM.
For this, we return to the specific case where $\rho = \ketbra{0}$ (though this also generalizes to arbitrary $\rho$).
As in Appendix \ref{sup-sec:dme-error}, we apply the unitary $e^{-i\SWAP\delta}$ to the joint state $\sigma \otimes \ketbra{0}$, and then randomly apply one of the unitaries $\{\pauli{\Id}, \pauli{Z}\}$ to the instruction qubit.
Then, it can be shown that the total state of instruction and target qubit is given by
\begin{align}
\label{Eq1}
&\frac{1}{2}\Big(e^{-i \SWAP\delta } \Big[\sigma\otimes \ketbra{0} \Big] e^{i \SWAP\delta }+(\pauli{\Id}\otimes\pauli{Z}) e^{-i  \SWAP \delta} \Big[\sigma\otimes \ketbra{0} \Big] e^{i  \SWAP \delta}(\pauli{\Id}\otimes\pauli{Z})  \Big)\nonumber\\
&=
     \underbrace{\mathcal{E}_{\delta\SWAP}^{\rho=\ketbra{0}}(\sigma)\otimes \ketbra{0}}_{\DME} - \underbrace{\sin^2(\delta) \ev{\sigma}{1}   \Big[\ketbra{0} \otimes \pauli{Z} \Big]}_{\text{\SQM{} error}}\ ,
\end{align}
where $\mathcal{E}_{\delta\SWAP}^{\rho=\ketbra{0}}(\sigma)$ is the quantum channel defined in Eq. \eqref{eq:e_dswap}.
Note that the first term,  $\mathcal{E}_{\delta\SWAP}^{\rho=\ketbra{0}}(\sigma)\otimes \ketbra{0}$ is exactly the desired state which can be used for the next round of \DME.
On the other hand, the second term $\sin^2(\delta)\ev{\sigma}{1}\Big[\ketbra{0}~\otimes~\pauli{Z}~\Big] $ can be treated as an error.
To find the contribution of this term in the total error, we use the fact that the trace-norm is non-increasing under any trace-preserving quantum operation $\mathcal{F}$: $\|\mathcal{F}(X)\|_\text{tr}\le \|X\|_\text{tr}$, where $\|\cdot \|_\text{tr}$ is trace norm, \textit{i.e.} sum of the absolute value of the eigenvalues of the operator.

 For the second term in Eq.~\eqref{Eq1} we have
 \beq
 \Big\| \sin^2(\delta)\ev{\sigma}{1} \Big[\ketbra{0}\otimes \pauli{Z}\Big]\Big\|_\text{tr}=2 \sin^2(\delta) \ev{\sigma}{1} \le 2\sin^2(\delta).
 \eeq
Therefore, the additional error introduced by each application of \SQM{} is bounded by $2\sin^2(\delta)$.

Repeating this process $N$ times, and using the triangle inequality for the trace norm, we find that the distance between the final total system state and the state produced by $\DME$ is bounded by $2N\sin^2(\delta)$.
Choosing $\delta=\theta/N$, we find that the overall additional error introduced by the use of \SQM{} is bounded by
\beq
2N\sin^2(\delta)=2N\sin^2\left(\frac{\theta}{N}\right) \le \frac{2\theta^2}{N}\ .
\label{eq-sup:SQM_error}
\eeq
The right hand side of Eq.~\eqref{eq-sup:SQM_error} is the \SQM-induced error contribution cited in the main text.

% -------------------------------------------------------------
%  IMPACT OF FINITE SQM
% -------------------------------------------------------------
\section{Quantifying the impacts of finite \SQM{} randomizations}\label{sup-sec:experimental_SQM_averaging}
To properly implement the probabilistic nature of the \SQM{} operation we instantiate each $\DME_2$ circuit a number of times. Consider as an example the $N=3$ version of the $\DME_2$ circuit from Figure~\figDMEfidelity,
\begin{equation}
\raisebox{-0.4cm}
{\Qcircuit @C=0.5em @R=0.5em{
\lstick{\sigma\vphantom{\sigma_\text{out}}} & \gate{\DME_2(|\!+\!i\rangle\langle +i|, 3, \pi/2)} & \rstick{\sigma_\text{out}} \qw
}}
\quad \quad \quad
\raisebox{-0.4cm}{$=$}
\quad \quad \quad \quad\quad
\Qcircuit @C=0.5em @R=2em{
 \lstick{\sigma\vphantom{\sigma_\text{out}}}				& \qw & \qswap		& \qw 				 & \qw				&  \qw & \qswap 		& \qw 				& \qw			& \qw & \qswap 		& \qw 				& \qw			& \rstick{\sigma_\text{out}}\qw\\
 \lstick{|\!+\!i\rangle\langle +i|}	& \qw & \qswap \qwx	& \qw \deltaExample	 & \gate{\SQM_y}	&  \qw & \qswap \qwx	& \qw \deltaExample	& \gate{\SQM_y}	& \qw & \qswap \qwx	& \qw \deltaExample	& \gate{\SQM_y} & \qw \\
 }
\end{equation}
In this case, each \SQM{} presents a random choice between applying $\RY{\pi}$ or $\Id$ at each occurence.
For an $N$ step $\DME_2$ there are $2^N$ configurations of \SQM{} gates.
In the experiment it is infeasible to sample all $2^N$ realizations, and instead we sample a smaller number, denoted $r$.
The circuits below show $r = 3$ random example realizations of the circuit,

\begin{equation}
\Qcircuit @C=0.5em @R=2em{
& \qw & \qswap		& \qw 				 & \qw				&  \qw & \qswap 		& \qw 				& \qw			& \qw & \qswap 		& \qw 				& \qw			& \qw\\
& \qw & \qswap \qwx	& \qw \deltaExample	 & \gate{\Id}	&  \qw & \qswap \qwx	& \qw \deltaExample	& \gate{\RY{\pi}}	& \qw & \qswap \qwx	& \qw \deltaExample	& \gate{\Id} & \qw \\
}
\quad\raisebox{-0.45cm}{$,$}\quad
\Qcircuit @C=0.5em @R=2em{
& \qw & \qswap		& \qw 				 & \qw				&  \qw & \qswap 		& \qw 				& \qw			& \qw & \qswap 		& \qw 				& \qw			& \qw\\
& \qw & \qswap \qwx	& \qw \deltaExample	 & \gate{\RY{\pi}}	&  \qw & \qswap \qwx	& \qw \deltaExample	& \gate{\RY{\pi}}	& \qw & \qswap \qwx	& \qw \deltaExample	& \gate{\RY{\pi}} & \qw \\
}
\quad\raisebox{-0.45cm}{$,$}\quad
\Qcircuit @C=0.5em @R=2em{
& \qw & \qswap		& \qw 				 & \qw				&  \qw & \qswap 		& \qw 				& \qw			& \qw & \qswap 		& \qw 				& \qw			& \qw\\
& \qw & \qswap \qwx	& \qw \deltaExample	 & \gate{\Id}	&  \qw & \qswap \qwx	& \qw \deltaExample	& \gate{\Id}	& \qw & \qswap \qwx	& \qw \deltaExample	& \gate{\RY{\pi}} & \qw \\
}
\end{equation}
In the experiment, a total $r_\SQM$ of circuits are executed, providing a sample from which we can extract average properties.
The generic process for extracting average properties over $r$ instantiations is sketched in Extended Data Fig. 7.
\begin{figure*}[!t]
\centering
	\includegraphics[width=1\columnwidth]{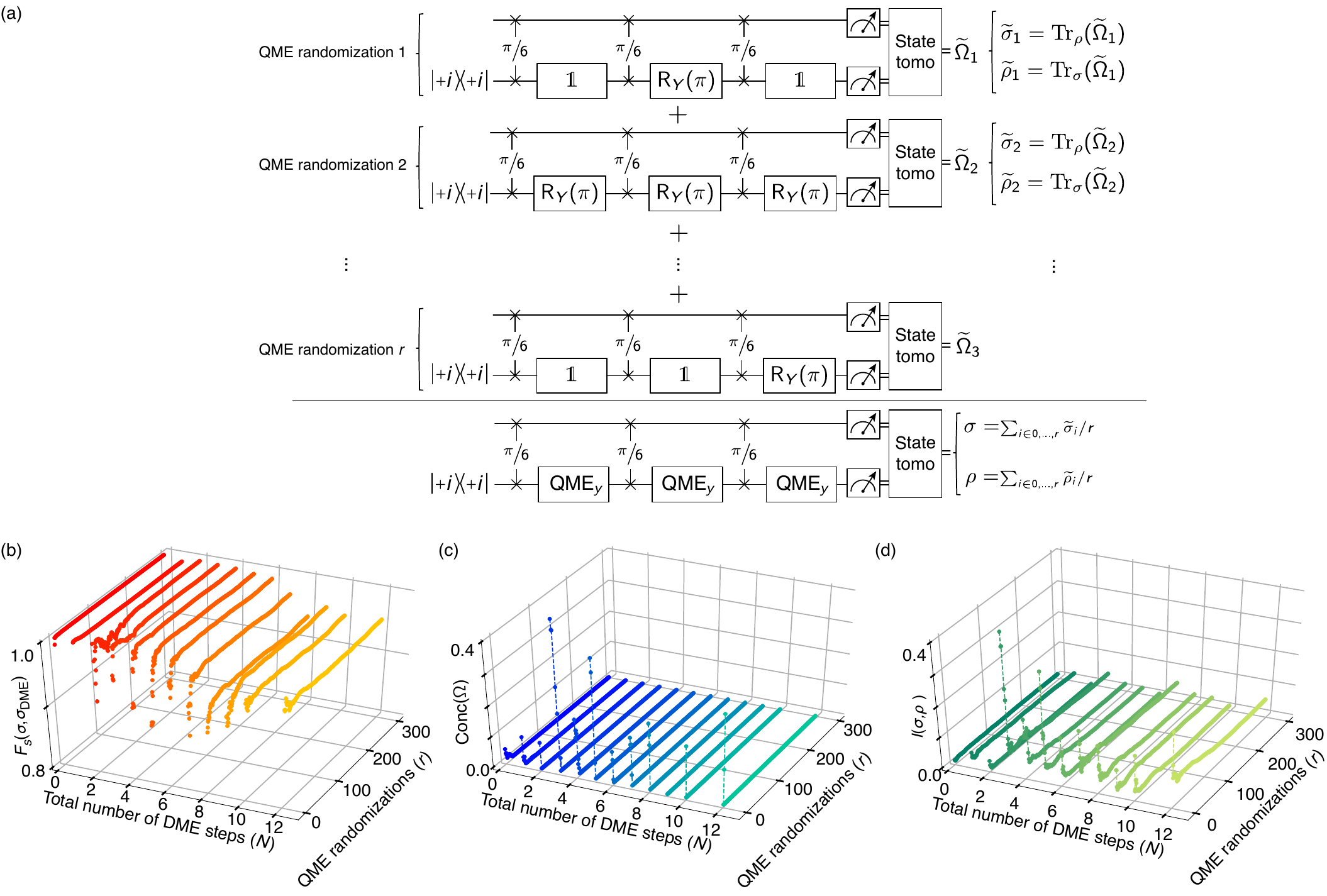}
		\caption{\footnotesize{
		(a) Schematic definition of experimental execution of a \DME{} protocol using \SQM{} operations (i.e., \DMEsqm).
		(b) The state fidelity between the measured output state and the result of ideal gates implementing \DME{}, as the number of \SQM{} randomizations are increased.
		(c) Concurrence in the two-qubit density matrix $\Omega$ (the combined state of the system), for increasing number of \SQM{} randomizations.
		(d) The mutual information between the two subsystems $\sigma$ and $\rho$, as more randomizations of \SQM{} are used.
		}}
	\label{fig-sup:sqm_randomizations}
\end{figure*}

From the datasets used in the main paper, we can also explore algorithmic behavior as the randomizations of \SQM{} increase toward the central limit.
In Extended Data Fig. 7b-d we plot three relevant figures of merit as a function of $r$ and $N$ for the $\theta = \pi$ dataset of Figure 3 in the main text.
Extended Data Fig. 7b shows the evolution of the state fidelity of the output state as a function of $r$.
For all values of $N$ we observe that after approximately $\sim$50 randomizations the effect of introducing more circuits with random choices of \SQM{} gates does not significantly alter the result.
Extended Data Fig. 7c shows the concurrence of the two-qubit density matrix, a measurement of bi-partite entanglement in the system \cite{wootters_entanglement_1998}.
After just a few randomizations $r> 10$, concurrence goes to zero, indicating that (quantum) correlations have been suppressed, as expected.
There may also be classical correlations between the $\sigma$ and $\rho$ subsystems.
In Extended Data Fig. 7d we therefore plot the mutual information $I(\sigma, \rho)$ between each subsystem, where
\begin{equation}
I_\Omega(\sigma,\rho) = S(\Tr_\sigma(\Omega))+S(\Tr_\rho(\Omega))-S(\Omega)
\end{equation}
is the mutual information, and $S(\Omega) = - \Tr\left(\Omega \ln \Omega\right)$ is the von Neumann entropy of the density matrix $\Omega$.
Here we again observe that after $r>10$ any correlations between the subsystems are effectively removed.

% -------------------------------------------------------------
%  BIBLIOGRAPHY
% -------------------------------------------------------------